\documentclass[]{aa}
\usepackage{graphicx}
\usepackage{txfonts}
\usepackage{graphics,graphicx,epstopdf,epsfig,url}
\usepackage{relsize}
\usepackage{dcolumn}
\usepackage{bm}
\usepackage[normalem]{ulem}
\usepackage{times}
\usepackage{natbib}
\usepackage{color}
\usepackage{multirow}
\usepackage{hyperref}
\usepackage{braket}
\usepackage[utf8]{inputenc}
\usepackage{xcolor}
\usepackage{hypcap}
\usepackage[autostyle]{csquotes}
\usepackage{textcomp }
\hypersetup{
    colorlinks=true, 
    citecolor=blue, 
    filecolor=black, 
    linkcolor=blue,
    urlcolor= black
}
\def\codename{\texttt{SCALAR} }
\def\codenamex{\texttt{SCALAR}}

\begin{document}

\title{\codenamex: an AMR code to simulate axion-like dark matter models}

 \author{Mattia Mina
          \inst{1}
          \and
          David F. Mota\inst{1}
          \and
          Hans A. Winther\inst{1,2}
          }

\institute{Institute of Theoretical Astrophysics, University of Oslo, 0315 Oslo, Norway\\
\email{mattia.mina@astro.uio.no}
\and
Institute of Cosmology \& Gravitation, University of Portsmouth, Portsmouth, Hampshire, PO1 3FX, UK}

\date{Date}

\abstract{
We present a new code, \codenamex, based on the high-resolution hydrodynamics and {\it N}-body code \texttt{RAMSES}, to solve the Schr\"odinger equation on adaptive refined meshes. The code is intended to be used to simulate axion or fuzzy dark matter models where the evolution of the dark matter component is determined by a coupled Schr\"odinger-Poisson equation, but it can also be used as a stand-alone solver for both linear and non-linear Schr\"odinger equations with any given external potential. This paper describes the numerical implementation of our solver and presents tests to demonstrate how accurately it operates. 
}

\keywords{methods: numerical, adaptive mesh refinement -- cosmology: axion-like dark matter, structure formation}

\maketitle

\section{Introduction}
The true nature of dark matter is not known. Weakly interacting massive particles (WIMP) are still considered one of the most likely candidates for cold dark matter (CDM), and several experiments are ongoing to try to detect such particles. These are, however, closing in on the neutrino floor where any signal would be drowned in the solar neutrino background \citep{2007PhRvD..76c3007M}. 

A promising alternative to WIMPs are ultra-light axions, fuzzy dark matter \citep{2016PhR...643....1M,2017PhRvD..95d3541H,PhysRevD.28.1243,PhysRevLett.64.1084,2000NewA....5..103G,Peebles_2000,49c0ac5f658e4b0693e5f368f6d94393,2003PhRvD..68b4023G,PhysRevD.93.025027,2006PhLB..642..192A, 2000PhRvL..85.1158H}, and superfluid dark matter \citep{2015PhRvD..92j3510B,2016arXiv160508443K,2015arXiv150703013K,2019JCAP...05..054S}. These models have distinct and observable signatures on the small scales of structure formation, they are also are able to solve some of the discrepancies observed in CDM simulations, such as the missing satellites problem \citep{1999ApJ...524L..19M,1999ApJ...522...82K}, the cusp-core problem \citep{2010AdAst2010E...5D,2018PhRvD..98h3027B,2019PhRvD..99j3020B} and the too-big-to-fail problem \citep{2011MNRAS.415L..40B}. These disparities could, however, have a solution within baryonic processes, which are usually not included in standard CDM simulations as shown in \citet{2012ApJ...744L...9M}, \citet{Brooks:2012ah}, \citet{2010Natur.463..203G}, \citet{2014ApJ...789L..17M}, \citet{2013MNRAS.429.3068T}, \citet{2012MNRAS.422.1231G}, \citet{2012MNRAS.421.3464P}, \citet{2016MNRAS.457.1931S}, and \citet{2017ApJ...850...97B}. To understand this better, one should ideally perform simulations including both of these components. 

In order to quantify the effects of axion-like dark matter models, one needs to either solve a Schr\"odinger-Poisson system or use the Mandelung formulation. The latter consists of a set of traditional hydrodynamics equations, with an additional pressure term which can be solved by using methods such as smoothed particle hydrodynamics (SPH), as proposed by \citet{2015PhRvD..91l3520M}. 
However, it is known that numerical methods based on the Madelung formulation of quantum mechanics are troublesome in regions around voids. This formulation, indeed, breaks down where the density approaches zero and at interference nodes, as the quantum pressure term can easily become singular \citep{2015PhRvD..91l3520M, 2019PhRvD..99f3509L}. 

The Schr\"odinger-Poisson system has several applications in cosmology. For instance, the six dimensional Vlasov equation describing collisionless self-gravitating matter is approximated by a Schr\"odinger-Poisson system for a complex wave-function in three dimensions. This was proposed as an alternative way for simulating CDM in \citet{1993ApJ...416L..71W}. It was later shown, by solving the Schr\"odinger-Poisson system and comparing it to the Vlasov solver \texttt{ColDICE} \citep{2016JCoPh.321..644S} in two dimensions, that one has excellent qualitative and quantitative agreement in the solution \citep{2017PhRvD..96l3532K}. A similar study is given in \citet{2018PhRvD..97h3519M}, where the system was solved using a spectral method, demonstrating that one recovers the classical behaviour in the limit where $\hbar \to 0$.

Unfortunately, the methods employed in the above mentioned papers, despite being very accurate, are too expensive to perform high-resolution simulations in three dimensions. The first cosmological, high-resolution, simulation of fuzzy dark matter in three dimensions was performed in \citet{2014NatPh..10..496S} using the code \texttt{GAMER} \citep{2010ApJS..186..457S,2018MNRAS.481.4815S}. There, an explicit method, similar to the one we present in this paper, was used. In \citet{PhysRevD.98.043509}, they used the classical wave-function to perform zoom-in simulations to study the formation and evolution of ultralight bosonic dark matter halos from realistic cosmological initial conditions and in \citet{2017MNRAS.471.4559M} they studied galaxy formation with Bose-Einstein condensate dark matter using a pseudo spectral method (see also \citet{2018JCAP...10..027E}). There have also been a handful of papers that have performed simulations more in line with the hydrodynamical formulation. In \citet{2018ApJ...853...51Z}, a new technique to discretise the quantum pressure is proposed and shown to reproduce the expected density profile of dark matter halos. In \citet{2018arXiv180108144N}, a module \texttt{AX-GADGET} for cosmological simulations using SPH inside the \texttt{P-GADGET3} code is presented. These methods do not solve for the wave-function, but they have the advantage of being much less expensive to run than a full wave-function solver like ours. There have also been simulations performed by using other numerical techniques in either two \citep{2017PhRvD..96l3532K} or three spatial dimensions \citep{2009ApJ...697..850W}.

In this paper we present \codename (Simulation Code for ultrA Light Axions in \texttt{RAMSES}): a general adaptive mesh refinement (AMR) code to solve the Schr\"odinger-Poisson system. Our implementation is in the hydrodynamics and {\it N}-body code \texttt{RAMSES} \citep{2002AA...385..337T}. 
The structure of the paper is as follows: in Section~\ref{sec:theory} we present the equations we are to solve, in Section~\ref{sec:numerical} we present the numerical implementation, in Section~\ref{sec:tests} we present tests of the code and in Section~\ref{sec:cosmo} we discuss possible cosmological applications before concluding in Section~\ref{sec:conc}.

\section{Theoretical model}\label{sec:theory}
A Bose-Einstein condensate (BEC) is a system of identical bosons, where a large fraction of particles occupies the lowest quantum energy state, or the ground state. This phenomenon typically takes place in gases, at very low temperatures or very high densities and it was observed for the first time in \citet{Anderson:1995gf} and \citet{PhysRevLett.75.3969}. In the condensate regime, these quantum systems behave as a macroscopic fluid and their peculiar features are a macroscopic manifestation of quantum effects. 

In general, when Bose-Einstein condensation occurs, thermal de-Broglie wavelengths of particles start to overlap, as they become grater than the mean inter-particle distance. At this point, a coherent state develops and the system behaves as a macroscopic fluid, where only binary collisions are relevant. The dynamics of BECs is complicated, due to the difficulty in modelling particle self-interactions. 

However, in the Hartree mean-field theory and in the limit of $T \to 0$, binary collisions are modelled via an effective potential and the whole quantum system can be described by a single-particle wave-function $\psi(\mathbf{x},t)$ obeying the non-linear Sch\"odinger equation:
\begin{align} \label{eq:GPE}
i\hbar\frac{\partial \psi}{\partial t} = \left [ -\frac{\hbar^2}{2m} \nabla^2 + g | \psi |^2 + mV_{\rm ext} \right ] \psi,
\end{align}
where $m$ is the mass of the boson and $g$ is the self-interaction coupling constant. Often, the trapping potential $V_{\rm ext}(\mathbf{x},t)$ is introduced by hand in order to model the presence of a trap, which is responsible for keeping particles confined.

The single-particle wave-function is normalised such that:
\begin{align}
\int | \psi |^2 ~ {\rm d}^3 x= N,
\end{align}
where $N$ is the total number of particles present in the system. As a consequence, the quantity $|\psi(\mathbf{x},t)|^2$ represents the number density of particles. 

An alternative description of the macroscopic fluid is provided by the so-called Madelung formulation of the Schr\"odinger equation. In this case, by expressing the single-particle wave-function in polar coordinates:
\begin{align} \label{eq:psipolar}
\psi = \sqrt{\frac{\rho}{m}} \exp \left ( \frac{i}{\hbar}\theta \right ),  
\end{align}
the dynamics of the system is described in terms of mass density and velocity, which are macroscopic physical quantities and they are respectively defined as:
\begin{align}
\rho(\mathbf{x},t) &= m |\psi(\mathbf{x},t)|^2, \label{eq:def_dens} \\
\vec{v}(\mathbf{x},t) &= \frac{1}{m} \vec{\nabla} \theta(\mathbf{x},t). \label{eq:def_vel}
\end{align}

Thus, the Schr\"odinger equation can be cast into the following system of equations:
\begin{align}
&\frac{\partial \rho}{\partial t} + \nabla \cdot (\rho \vec{v}) = 0, \label{eq:madelung1} \\
&\frac{\partial \mathbf{v}}{\partial t} + (\vec{v}\cdot \nabla)\vec{v} = -  \nabla \left ( V_{\rm ext} + \frac{g}{m^2}\rho + Q \right ), \label{eq:madelung2}
\end{align}
which are known as the Madelung or quantum Euler equations. We recognise Eq.~\eqref{eq:madelung1} as a continuity equation which expresses conservation of mass. Although the second Madelung equation, Eq.~\eqref{eq:madelung2}, expresses conservation of momentum, it is not the same as the classical momentum equation, as it contains an additional term $Q$, which is called quantum pressure and it is defined as:
 \begin{align} \label{eq:QP}
Q \equiv -\frac{\hbar^2}{2m^2}\frac{\nabla^2\sqrt{\rho}}{\sqrt{\rho}}.
\end{align}
The quantum pressure is a macroscopic manifestation of quantum effects and it is characteristic of Bose-Einstein condensates.

In this formulation, by defining the velocity as in Eq.~\eqref{eq:def_vel}, we are intrinsically assuming that the fluid is irrotational, since:
\begin{align}
\vec{\nabla} \times \vec{v} = \vec{\nabla} \times \vec{\nabla} \theta = 0.
\end{align}
However, during the evolution of the wave-function, the phase can develop discontinuities of multiples of $2\pi\hbar$ and its gradient can subsequently generate vorticity in the field, as shown in \citet{2014PhRvD..90b3517U,2019PhRvD..99h3524U}.

In cosmology, these kinds of systems can be used to model the dark matter contribution to the energy budget of the Universe. In particular, in the last few decades, models where dark matter is a light boson, such as ultra-light axions or fuzzy dark matter, have received a lot of attention. Due to the small mass of these bosons, macroscopic quantum effects manifest at astronomically relevant scales. In these alternative dark matter models, new signatures are expected within the structure formation process at highly non-linear scales and, therefore, numerical simulations are required in order to explore these scenarios.

Here, the dynamics of dark matter is also described by a system of identical bosons gravitationally bounded. Therefore, the governing equation is a non-linear Schr\"odinger equation, Eq.~\eqref{eq:GPE}, where the external potential is replaced by the gravitational potential. In this class of alternative dark matter models, self-interactions between bosons are often neglected, as the coupling constant $g$ is usually parametrically small. The resulting system of equations describing the dynamics of the dark matter fluid is called Schr\"odinger-Poisson system and, for an expanding Universe, it reads:
\begin{align}\label{eq:axioncosmo}
&i\hbar \left ( \dfrac{\partial \psi}{\partial t} + \dfrac{3}{2}H\psi \right ) = \left ( -\dfrac{\hbar^2}{2ma^2}\nabla^2 + m_a\Phi \right ) \psi, \\
&\nabla^2 \Phi = 4\pi G a^2 \left ( |\psi|^2-|\psi(a)|^2 \right ),
\end{align}
where $a$ is the scale-factor of the Universe, $H \equiv d\log(a)/dt$ is the Hubble rate of expansion, and $\Phi$ is the gravitational potential. With a change of variables $\psi \to a^{3/2}\psi$, the non-linear Schr\"odinger equation above takes on the form of Eq.~\eqref{eq:GPE}.

\section{Numerical methods}\label{sec:numerical}
In this section we provide a brief overview of the code \texttt{RAMSES} and the AMR technique. Then, we discuss in details the numerical aspects of the algorithm we implemented in order to solve the non-linear Schr\"odinger equation. Throughout this section, the dimensionality of the problem is denoted by $\rm{dim}$ and it can be 1, 2 or 3.

\subsection{Overview of \texttt{RAMSES}}

The \texttt{RAMSES} code was originally designed for cosmological simulations of structure formation and subsequently extended to astrophysical applications. It consists of an {\it N}-body particle mesh (PM) code, which solves the gravitational dynamics of a set of macroparticles, sampling the phase space distribution of the dark matter component in the Universe. Through PM algorithms, the mass of each macroparticle is interpolated on a grid and the Poisson equation is solved in order to compute the gravitational potential. Thus, the gravitational force acting on each element of the system and the new phase space position of each macroparticle are computed by solving the corresponding {\it N}-body equation with a leapfrog scheme. In addition, \texttt{RAMSES} can solve the dynamics of the baryonic component present in the Universe. In this case, the grid is also used to sample gas parcels and the evolution of the system is described by the equations of hydrodynamics, which are solved by means of a Godunov scheme. For this purpose, Riemann solvers can be used for computing fluxes of conserved physical quantities among cells.

The \texttt{RAMSES} code implements an AMR strategy, where a hierarchy of nested grids is created in order to increase the local resolution according to a set of refinement criteria. In this way, \texttt{RAMSES} can solve accurately gas dynamics and gravitational potential only where more precision is actually needed. This approach reduces consistently the amount of memory needed in cosmological and hydrodynamical simulations, compared to the case where a uniform high-resolution grid is used.

In \codenamex, we rely on the efficient AMR implementation of \texttt{RAMSES}. In order to solve the dynamics of our theoretical model, the single-particle wave-function is sampled by using the original grid allocated by \texttt{RAMSES} for the Poisson and hydrodynamics equations. Also in this case, the AMR approach provides the opportunity to solve the Schr\"odinger equation with higher resolution only where features of the wave-function are more demanding. 

\subsection{Adaptive mesh refinement}
The basic unit of the AMR hierarchy is an oct, which consists of a set of $2^{\rm dim}$ cells. At each level in the AMR hierarchy, a grid is a collection of octs with the same resolution. The grid with the coarsest resolution is called domain grid and it covers the whole computational domain. During the evolution of the physical system, when the solution starts to develop features and its tracking requires higher resolution, any cell at a given level can be split into a child oct, with double the resolution of the parent cell.

At each time-step, the AMR structure is modified according to a set of refinement criteria. First, for a generic level of refinement $\ell$, a refinement map is created by marking all those cells satisfying at least one refinement criterion. Also cells violating the strict refinement rule are marked for refinement, in order to guarantee that each child oct at level $\ell+1$ is surrounded by, at least, $3^{\rm dim}-1$ neighbours at the coarser level. However, if a given cell at level $\ell$ does not satisfy any refinement criteria anymore, it is marked for de-refinement and subsequently its child octs are destroyed. Then, a new child oct is created at level $\ell+1$ for each marked cell and all the relevant physical quantities are interpolated from level $\ell$. Coarse-fine data interpolation, in an AMR context, is often called prolongation and it can be done by using any of the interpolation schemes which are described in the section below.

When computing the refinement map, physical quantities can fluctuate around the refinement threshold in subsequent time-steps. This means that, some cells at the boundary of a fine resolution region can be refined and de-refined many subsequent times. In this case, the refinement map tends to be quite noisy, since each interpolation operation where the solution is characterised by strong gradients introduces numerical noise in the solution of the non-linear Schr\"odinger equation. For this reason, once the refinement map is computed according to a given set of refinement criteria, a mesh smoothing operator is applied. For this purpose, a cubic buffer of $n_{\rm expand}$ cells around the computed map is additionally marked for refinement. In this way, even if cells octs are created and destroyed at coarse-fine boundaries, the interpolation procedure is applied in a regions where the wave-function is smoother and, thus, it introduces a lower level of numerical noise.

\subsection{The Schr\"odinger equation}
The \codename code evolves the solution of the non-linear Schr\"odinger equation by using a Taylor's method, similar to the one designed in \texttt{GAMER}. 
Given the wave-function $\psi(\mathbf{x},t_0)$, the formal solution of the non-linear Schr\"odinger equation, Eq.~\eqref{eq:GPE}, at time $t_1 = t_0 +  \Delta t$ reads:
\begin{align}\label{eq:formal_sol}
\psi(\mathbf{x},t_1) = \hat{U}(t_1,t_0) \psi (\mathbf{x},t_0),
\end{align}
where $\hat{U}(t_1,t_0)$ is the time evolution operator and it maps the solution of the Schr\"odinger equation at two different times. In the general case, the time evolution operator is defined as:
\begin{align} \label{eq:time_evo}
\hat{U}(t_1,t_0) = \exp \left ( -\frac{i}{\hbar} \int_{t_0}^{t_1} \hat{H}(\mathbf{x},t') ~ {\rm d}t' \right ),
\end{align}
where $\hat{H}(\mathbf{x},t)$ denotes the Hamiltonian of the system. The operator $\hat{U}(t_1,t_0)$ has the main following properties:
\begin{itemize}
\item 
$\begin{aligned}[t]
\hat{U}(t,t) = 1,
\end{aligned}$
\item 
$\begin{aligned}[t]
\hat{U}(t_1,t_2) ~ \hat{U}(t_2,t_3) = \hat{U}(t_1,t_3),
\end{aligned}$
\item 
$\begin{aligned}[t]
\hat{U}(t_1,t_2) = \hat{U}^{\dag}(t_2,t_1) = \hat{U}^{-1}(t_2,t_1).
\end{aligned}$
\end{itemize} 
In the limit of $ \Delta t \ll 1$, then the following approximation holds:
\begin{align}
 \int_{t_0}^{t_1} \hat{H}(\mathbf{x},t') ~ {\rm d}t' \approx \hat{H}(\mathbf{x},t_0) ~ \Delta t,
\end{align}
and, therefore, Eq.~\eqref{eq:time_evo} can be approximated as:
\begin{align} \label{eq:time_evo_discr}
\hat{U}(t_1,t_0) \approx  \exp \left ( {-\frac{i}{\hbar} \hat{H}(\mathbf{x},t_0) \Delta t} \right ).
\end{align}
In the general case, the Hamiltonian $\hat{H}(\mathbf{x},t)$ contains different contributions to the total energy of the system. In particular, we can express $\hat{H}(\mathbf{x},t)$ as a sum of contributions describing kinetic and potential energies. Here, we denote these two operators respectively as $\hat{K}(\mathbf{x},t)$ and $\hat{W}(\mathbf{x},t)$, and they are defined as: 
\begin{align}
&\hat{K}(\mathbf{x},t) \equiv -\frac{\hbar^2}{2m}\nabla^2, \\
&\hat{W}(\mathbf{x},t) \equiv m \left(V(\mathbf{x},t) + \frac{g}{m}|\psi(\mathbf{x},t)|^2\right).
\end{align}
By means of the Lie-Trotter formula \citep{10.2307/2033649}, the time evolution operator can be split as well:
\begin{align} \label{eq:time_evo_split}
\hat{U}(t_1,t_0) \approx \exp \left ( {-\frac{i}{\hbar} \hat{W}(\mathbf{x},t_0) \Delta t} \right ) \exp \left ( {-\frac{i}{\hbar} \hat{K}(\mathbf{x},t_0) \Delta t} \right ).
\end{align}
As a consequence, the formal solution of the Schr\"odinger equation can be written as:
\begin{align}
\psi (\mathbf{x},t_1) = &\exp \left ( {-\frac{i}{\hbar} \hat{W}(\mathbf{x},t_0) \Delta t} \right ) \exp \left ( {-\frac{i}{\hbar} \hat{K}(\mathbf{x},t_0) \Delta t} \right ) \psi(\mathbf{x},t_0).
\end{align}
In \codename, the two contributions to the time evolution operator are applied separately. First, the \textquotesingle drift\textquotesingle ~due to the kinetic part of the Hamiltonian is approximated via Taylor expansion (here for ${{\rm dim}} = 3$):
\begin{align} \label{eq:taylor_K}
\bar{\psi}_{i,j,k}^{n+1} &= \exp \left ( {-\frac{i}{\hbar} \hat{K}(\mathbf{x},t_0) \Delta t} \right ) \psi^n_{i,j,k} \nonumber \\
&= \left [ \sum^{\infty}_{N=0} \frac{1}{N!} \left ( -\frac{i}{\hbar} \hat{K}(\mathbf{x},t_0) \Delta t \right )^{N} \right ] \psi^n_{i,j,k} \nonumber \\
&= \left [ 1 + \left(\frac{i\hbar \Delta t}{2m}\nabla^2\right) + \frac{1}{2}\left(\frac{i\hbar \Delta t}{2m}\nabla^2\right)^2 + \ldots\right] \psi^n_{i,j,k},
\end{align}
where, for a generic operator $\hat{O}$, the notation $\hat{O}^{N}$ denotes $N$ consecutive applications of the same operator. In the \codename code, the Taylor expansion is performed up to $\mathcal{O}( \Delta t^3)$, which is the minimum required by the stability analysis of the numerical scheme. Furthermore, the laplacian operator is discretised by a standard second-order finite difference formula:
\begin{align}
\nabla^2 \psi_{i,j,k}^n = &\frac{\psi_{i+1,j,k}^n+\psi_{i-1,j,k}^n-2\psi^n_{i,j,k}}{\Delta x^2}+ \nonumber\\
&\frac{\psi_{i,j+1,k}^n+\psi_{i,j-1,k}^n-2\psi^n_{i,j,k}}{\Delta x^2}+ \nonumber\\
&\frac{\psi_{i,j,k+1}^n+\psi_{i,j,k-1}^n-2\psi^n_{i,j,k}}{\Delta x^2}.
\end{align}
Then, the \textquotesingle kick\textquotesingle ~due to the potential is computed and the wave-function at the new time-step reads:
\begin{align} \label{eq:exp_W}
\psi^{n+1}_{i,j,k} = \exp \left ( {-\frac{i}{\hbar} \hat{W}^n_{i,j,k} \Delta t} \right ) \bar{\psi}_{i,j,k}.
\end{align}
Here, the advantage of the Lie-Trotter splitting is clear: while the kinetic contribution to the time evolution operator needs Taylor expansion in order to be applied, the potential contribution only provides a phase rotation of the wave-function and it can be computed exactly. 

Once the new wave-function is computed, the new mass density is computed according to:
\begin{align}
\rho_{i,j,k}^{n+1} \big |_S = m \left | \psi_{i,j,k}^{n+1} \right | ^ 2.
\end{align}

\subsection{The continuity equation}
In quantum mechanics, the time evolution operator is unitary, as expressed by its properties. This means that the mass density carried by the wave-function is conserved. This is true also if we consider separately the two contributions to the Hamiltonian. However, the Taylor expansion, Eq.~\eqref{eq:taylor_K}, breaks the unitarity of the time evolution operator. Therefore, in order to improve the conservation properties of our main numerical scheme, we implement a secondary solver for the continuity equation associated to the non-linear Schr\"odinger equation. 

Eq.~\eqref{eq:madelung1} can be written in its conservative form:
\begin{align} \label{eq:cont_eqn_cons}
\frac{\partial \rho}{\partial t} + \mathbf{\nabla} \cdot \mathbf{j} = 0,
\end{align}
where $\rho(\mathbf{x},t)$ represents the mass density. Here, the quantity $\mathbf{j}(\mathbf{x},t)$ is the associated density current, or flux, and it is defined as:
\begin{align} \label{eq:fluxes}
\mathbf{j} \equiv -i\frac{\hbar}{2m} \left ( \psi^{*} \mathbf{\nabla} \psi - \psi \mathbf{\nabla} \psi^{*} \right ).
\end{align} 
By explicitly considering real and imaginary part of the wave-function, the density current can also be expresses as: 
\begin{align}\label{eq:fluxes2}
\mathbf{j} = \frac{\hbar}{m} \left ( \Re[\psi] \nabla \Im[\psi] - \Im[\psi] \nabla \Re[\psi] \right ).
\end{align}
In \codenamex, Eq.~\eqref{eq:cont_eqn_cons} is discretised by using a first-order Godunov scheme:
\begin{align} \label{eq:new_dens}
\frac{\rho_{i,j,k}^{n+1} - \rho_{i,j,k}^{n}}{\Delta t} &+ \frac{ \left (j_{i+\frac{1}{2},j,k}^{n+\frac{1}{2}} - j_{i-\frac{1}{2},j,k}^{n+\frac{1}{2}} \right )}{\Delta x} \nonumber\\
&+ \frac{\left (j_{i,j+\frac{1}{2},k}^{n+\frac{1}{2}} - j_{i,j-\frac{1}{2},k}^{n+\frac{1}{2}} \right )}{\Delta x} \nonumber\\
&+ \frac{\left (j_{i,j,k+\frac{1}{2}}^{n+\frac{1}{2}} - j_{i,j,k-\frac{1}{2}}^{n+\frac{1}{2}} \right ) }{\Delta x} = 0,
\end{align}
where the time-centered fluxes are computed at cell interfaces. In order to compute the time-centered fluxes at cell interfaces, the wave-function is first computed at half time-step, by advancing the solution of $0.5 \Delta t$. Then, the wave-function at cell interfaces is estimated in each dimension by linear interpolation:
\begin{align}\label{eq:lin_int_psi}
\psi_{i+\frac{1}{2},j,k}^{n+\frac{1}{2}} &= \frac{\psi_{i,j,k}^{n+\frac{1}{2}}+\psi_{i+1,j,k}^{n+\frac{1}{2}}}{2}, \nonumber \\
\psi_{i,j+\frac{1}{2},k}^{n+\frac{1}{2}} &= \frac{\psi_{i,j,k}^{n+\frac{1}{2}}+\psi_{i,j+1,k}^{n+\frac{1}{2}}}{2}, \\
\psi_{i,j,k+\frac{1}{2}}^{n+\frac{1}{2}} &= \frac{\psi_{i,j,k}^{n+\frac{1}{2}}+\psi_{i,j,k+1}^{n+\frac{1}{2}}}{2}. \nonumber
\end{align}
Its gradient, instead, is computed in each dimension by means of the first-order finite difference formula and it reads:
\begin{align}\label{eq:lin_int_nabla_psi}
\nabla\psi_{i+\frac{1}{2},j,k}^{n+\frac{1}{2}} &= \frac{\psi_{i+1,j,k}^{n+\frac{1}{2}}-\psi_{i,j,k}^{n+\frac{1}{2}}}{\Delta x}, \nonumber \\
\nabla\psi_{i,j+\frac{1}{2},k}^{n+\frac{1}{2}} &= \frac{\psi_{i,j+1,k}^{n+\frac{1}{2}}-\psi_{i,j,k}^{n+\frac{1}{2}}}{\Delta x}, \\
\nabla\psi_{i,j,k+\frac{1}{2}}^{n+\frac{1}{2}} &= \frac{\psi_{i,j,k+1}^{n+\frac{1}{2}}-\psi_{i,j,k}^{n+\frac{1}{2}}}{\Delta x}. \nonumber
\end{align}
Thus, the time-centered density current at cell interfaces is computed by means of Eq.~\eqref{eq:fluxes2}.

However, this solver is not used to explicitly advance in time the mass density, but only to enforce the conservation of mass. Indeed, by denoting $\rho_{i,j,k}^{n+1} \big |_S$ and $\rho_{i,j,k}^{n+1} \big |_C$ the new mass densities computed by the main and the secondary solvers respectively, a correcting factor is computed as follows:
\begin{align} \label{eq:r}
R = \frac{\rho^{n+1}_{i,j,k} \big |_C}{\rho^{n+1}_{i,j,k} \big |_S},
\end{align}
which is used to rescale the wave-function.

Although by solving the continuity equation on top of the Schr\"odinger equation sensibly improves the conservation properties of the algorithm, this process does not ensure perfect conservation of mass. Since we truncate at the third order in time the Taylor expansion of the kinetic operator, Eq. \eqref{eq:taylor_K}, the kinetic solver introduces truncation errors in the solution of the Schr\"odinger equation. However, by rescaling the wave-function with the correct amplitude computed from the continuity equation, we are able to significantly reduce truncation errors, as shown by the improvement of several orders of magnitude on the conservation properties of \codename in the tests we present in Section \ref{sec:tests}. 
Unfortunately, the rescaling procedure is subject to accumulation of round-off errors, leading to an evolution of the error in the conservation of mass itself. However, the round-off errors are of the order of machine precision, in contrast to the much higher amplitude of the truncation errors introduced by the kinetic solver. This solution was already adopted in the \texttt{GAMER} code \citep{2014NatPh..10..496S}.

\subsection{The solver}
\codename solves the Schr\"odinger equation from the coarser to the finer level in the AMR hierarchy. For a generic refinement level $\ell$, the optimal time-step is chosen as:
\begin{align} \label{eq:CFL}
 \Delta t = \min\left[C_K \cdot \frac{\sqrt{3}}{2\hbar}m(\Delta x)^2,~~C_W \cdot \frac{2\pi \hbar}{m|V_{\rm max}|}\right],
\end{align}
where $|V_{\rm max}|$ denotes the maximum absolute value of the effective potential $V + \frac{g}{m}|\psi|^2$. Here, $C_K$ and $C_W$ are Courant factors which are required to be smaller than one. The first term in the square brackets is determined by the Von Neumann stability analysis of the kinetic part of the solver. The second term, instead, requires that the phase of the wave-function does not rotate by a bigger angle than $2\pi C_W$ within a time-step. 
In general, the Courant factors $C_K$ and $C_W$ are chosen empirically, depending on the characteristics of the physical system that one aims to model. In our case, we set $C_K = C_W = 0.2$ since it provides a good accuracy on the solution of the non-linear Schr\"odinger equation, without sacrificing too much computation time. In Appendix~\ref{sect:app_stability} we provide a detailed discussion of the Von Neumann stability analysis of the numerical scheme. 

In the original \texttt{RAMSES} code two different options are available regarding the choice of the time-step: a single or an adaptive time-step. While the former consists in using the same time-step for all refinement levels and it is determined by the finest level in the AMR hierarchy, the latter allows to use smaller time-steps for finer refinement levels. However, in case of adaptive time-step, for each coarse time-step at level $\ell$ it is possible to perform only two fine time-steps at level $\ell+1$. 
In \codenamex, an additional option is available: a flexible time-step, where for each coarse time-step at level $\ell$, the number of fine steps at level $\ell+1$ is flexible and it is determined by a level dependent Courant-Friedrichs-Lewy (CFL) condition. From Eq. \eqref{eq:CFL}, when the optimal time-step is chosen by the kinetic CFL condition, the time-step scales with the grid size as $\Delta t \propto \Delta x^2$, which represents a stricter condition than the usual case of hydrodynamics equations. Therefore, a flexible time-step can reduce significantly the total amount of coarse time-step in a simulation.

\begin{figure}
\centering
\includegraphics[width=0.9\columnwidth]{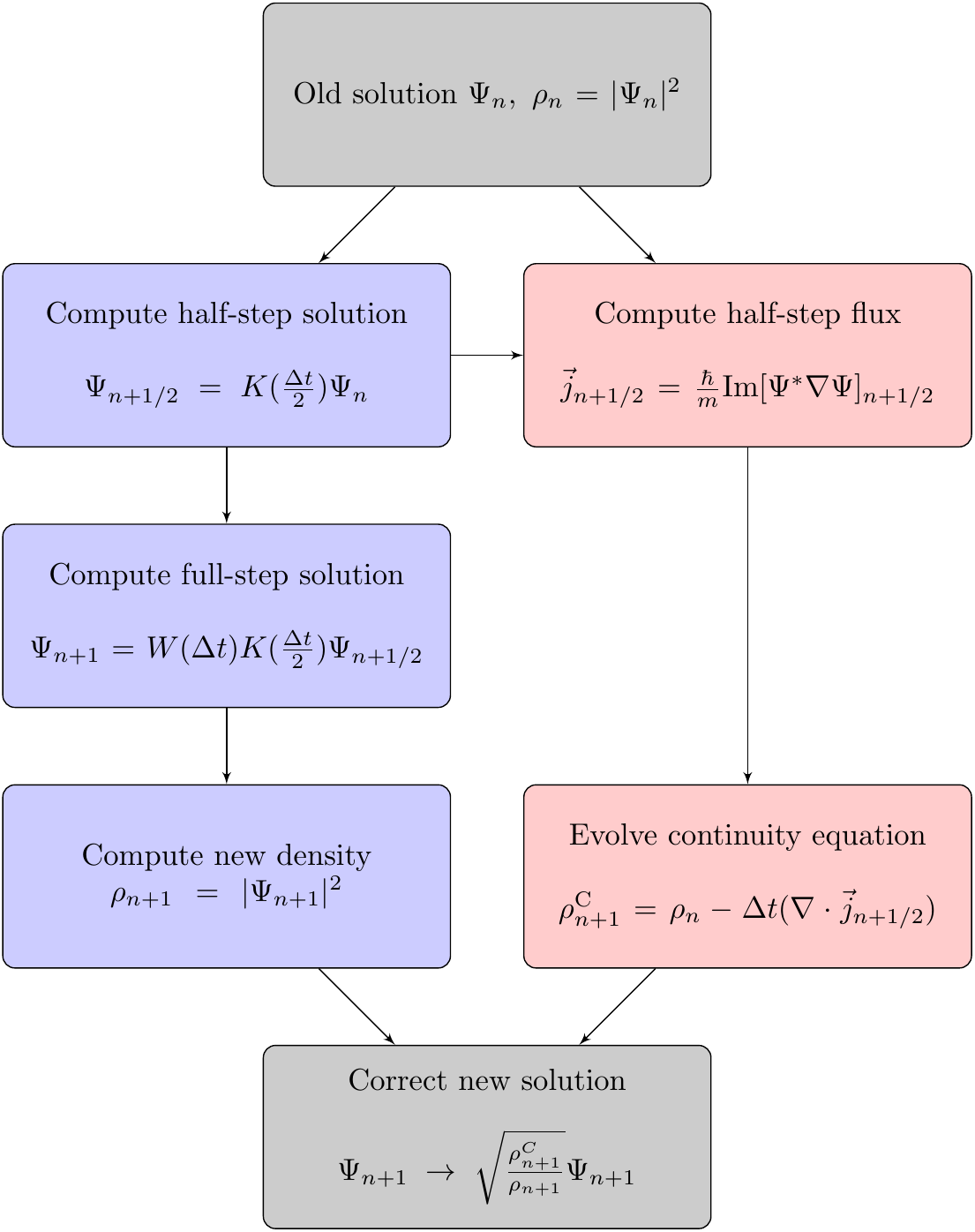}
\caption{Flowchart of the numerical algorithm implemented to solve the Schr\"odinger-Poisson system. The kinetic solver is shown in blue (left) and the continuity solver is shown in red (right).}
\label{fig:flowchart}
\end{figure}

Within a generic level of refinement $\ell$, \codename solves the non-linear Schr\"odinger equation for each oct separately. Thus, in order to advance the solution over a time-step, the solver proceeds as follows:
\begin{enumerate}
\item For a given oct, a cubic buffer of neighbours cells is collected. The equations are actually solved only for the central oct, while the buffer cells are used to compute laplacians at each order of the Taylor expansion. If the central oct lies next to the coarse-fine boundary, the wave-function is interpolated into ghost cells from level $\ell-1$.
\item The kinetic solver evolves the wave-function at the new time-step by means of Eq.~\eqref{eq:taylor_K}. First, by advancing the solution by $0.5\Delta t$, the half time-step solution $\bar{\psi}_{i,j,k}^{n+1/2}$ is estimated from $\psi_{i,j,k}^{n}$ and it is used later in order to compute the mass density currents. Then, the full-step solution $\bar{\psi}_{i,j,k}^{n+1}$ and the new mass density $\rho_{i,j,k}^{n+1} \big |_S$ are computed.
\item The wave-function at half time-step $\bar{\psi}_{i,j,k}^{n+1/2}$ is interpolated at cell interfaces by using Eqs.~\eqref{eq:lin_int_psi}--\eqref{eq:lin_int_nabla_psi} and the time-centered density currents are computed by means of Eq.~\eqref{eq:fluxes2}.
\item The continuity equation is solved and the new mass density $\rho_{i,j,k}^{n+1} \big |_C$ is computed via Eq.~\eqref{eq:new_dens}.
\item The rescaling factor given by Eq.~\eqref{eq:r} is computed and $\bar{\psi}_{i,j,k}^{n+1}$ is rescaled in order to preserve mass conservation.
\item If the given oct lies next to the coarse-fine boundary, the estimated flux is stored for the subsequent reflux operation.
\item Finally, the phase rotation due to the potential is computed by applying Eq.~\eqref{eq:exp_W} to $\bar{\psi}_{i,j,k}^{n+1}$, and $\psi_{i,j,k}^{n+1}$ is evaluated.
\end{enumerate}
All steps, except the first and the last ones, are performed separately for each physical dimension. This procedure is called dimensional splitting and it reduces a $N$-dimensional problem into a system of $N$ one dimensional problems. It has the advantage of relaxing the CFL condition of the solver and, therefore, it allows bigger time-steps. A flowchart of our solver is shown in Fig.~(\ref{fig:flowchart}).

\subsection{Refinement strategy} 
In \codenamex, we implement the same \textquotesingle quasi-Lagrangian\textquotesingle ~approach as \texttt{RAMSES} uses for hydrodynamics: when the total mass of a given cell exceeds a given threshold, the cell is marked for refinement. The level dependent density threshold is defined as:
\begin{align}\label{eq:masscrit}
\rho_{\ell} = \frac{M_c}{(\Delta x_{\ell})^{\rm dim}},
\end{align}
where $M_c$ corresponds to the maximum mass allowed per cell.

In addition, following \citet{2014NatPh..10..496S} and the \texttt{FLASH} code \citep{Fryxell2000}, we implement support for the invariant version of the L\"ohner error estimator. It is based on the second derivative of a given physical quantity, normalised by the first derivative. Considering a generic physical quantity $f$, the error estimator $E_L$ reads:
\begin{align} \label{eq:lohner}
E_L = \left \{ \frac{ \sum_{ij} \left ( \dfrac{\partial^2 f}{\partial x_i \partial x_j} \right )^2 }{ 
\sum _{ij} \left [ 
\dfrac{1}{2\Delta x_j} \left ( \left | \dfrac{\partial f}{\partial x_i} \right |_{i+1} + \left | \dfrac{\partial f}{\partial x_i} \right |_{i-1} \right ) + \xi \dfrac{ |\bar{f}_{ij}| }{ \Delta x_i \Delta x_j }
 \right ]^2
} \right \}^{1/2},
\end{align}
where the indices $i$,$j$ run over each physical dimensions. Small fluctuations of the physical quantity $f$ are filtered out due to the presence of the second term at denominator. The quantity $|\bar{f}_{ij}|$ is an average of $f$ over dimensions $i$,$j$ and $\xi$ is a small constant. This error estimator is dimensionless and therefore it can be applied to any physical quantity. Furthermore, in Eq. \eqref{eq:lohner}, $E_L$ is bounded in the interval $[0,1]$. In \codenamex, we apply the L\"ohner error estimator separately to ${\Re[\psi]}$ and ${\Im[\psi]}$. Then, the final estimation of the error on the wave-function is given by:
\begin{align}
E_L = \sqrt{ \left (E_L^{ \Re } \right )^2 + \left ( E_L^{ \Im } \right )^2 },
\end{align}
and if it exceeds a user-defined threshold, the cell is marked for refinement. 
This threshold can be chosen empirically, depending on the features one needs to resolve in the solution of the governing equations. Although it is currently implemented in \codenamex, we did not employed the L\"ohner error estimator in the test we present here, in Section \ref{sec:tests}. In general, while testing separately the implementation of the L\"ohner error estimator, we find that a value of $E_L = 0.7$ provides a good balance between computational cost and accuracy for the solution of the non-linear Schr\"odinger equation.

\subsection{Spatial and temporal interpolation.} 
In \codenamex, interpolation is required when a generic level $\ell$ in the AMR hierarchy is refined and new child octs are created at level $\ell+1$,
or when, during the solving process, boundary conditions need to be specified for a fine-grid patch and ghost cells are created.
In both cases, the wave-function at the coarse level $\ell$ is interpolated down to level $\ell +1$. In order to solve the equations of motion, when the laplacian operator is applied, any discontinuity in the second derivative of the wave-function introduces an error, which propagates into the solution of the non-linear Schr\"odinger equation and it can destroy the wave-function. Therefore, high-order interpolation schemes are implemented in order to keep the wave-function as smooth as possible.

In particular, in \codenamex, we implement two high-order interpolation schemes. In both cases, the interpolating function is a fourth-order polynomial, but the coefficients of the polynomials are chosen in different ways. In one case, Lagrange basis polynomials are computed in order to set the coefficients, resulting in a fourth-order Lagrange interpolation scheme. In the second case, fourth-order conservative interpolation is performed and the coefficients of the interpolation are set by requiring that cell averages of the interpolated quantities are preserved. In case of adaptive time integration, linear temporal interpolation can also be applied when computing boundary conditions for a fine level patch, since coarse-grid and fine-grid wave-functions can be discretised at different times.

Furthermore, the interpolation can be performed on two different sets of variables: the original set of variables ${\Re[\psi]}$ and ${\Im[\psi]}$, or derived variables $m|\psi|^2$ and $\rm{Arg}[\psi]$, corresponding to mass density and phase of the wave-function. The interpolation schemes and the set of variables used for the interpolation process can be specified by the user in the parameter file.

Further details on the interpolation schemes can be found in Appendix~\ref{sect:app_prolong}.

\subsection{Artificial viscosity}

In the tests shown in the upcoming sections, when they were done at the domain level only, the solution of the non-linear Schr\"odinger equation remains stable for as long as we could run \codenamex. However, when refinements were included, the solver had the tendency to develop spurious high-frequency waves at coarse-fine boundary, even after improving the order of accuracy of interpolation schemes. In order to artificially dump spurious oscillations, we introduced an empirical viscosity term in the non-linear Schr\"odinger equation. Thus, incorporating the viscosity term, the non-linear Schr\"odinger equation, Eq.~\eqref{eq:GPE}, is replaced by: 
\begin{align}
i\hbar\frac{\partial \psi(\mathbf{x},t)}{\partial t} = \left [ -\frac{\hbar}{2m} \left( 1- i \epsilon \right) \nabla^2 + g | \psi(\mathbf{x},t) |^2 + mV_{\rm ext}(\mathbf{x},t) \right ] \psi(\mathbf{x},t),
\end{align}
where the constant $\epsilon>0$ quantifies the strength of the damping term. For example, if we consider a single plane-wave:
\begin{align}
\psi \propto \exp(i\omega t - ikx),
\end{align}
the viscosity term acts in a similar way as a Gaussian filter, by dumping the wave-function by a factor of:
\begin{align}
\exp \left ( -\frac{k^2 \epsilon t}{2m} \right ).
\end{align}
This means that, including the artificial viscosity, the wave-function is simply replaced by:
\begin{align}
\psi \to \psi \exp \left ( -\frac{k^2 \epsilon t}{2m} \right ),
\end{align}
In this way, the filter leaves untouched physical low frequency modes in the wave-function, while smoothing the spurious numerical oscillations.
In general, an artificial viscosity term would affect mass conservation. However, by solving the continuity equation on top of the non-linear Schr\"odinger equation, mass conservation is enforced and the artificial viscosity simply acts as a viscous force. Indeed, if we consider the Madelung formulation of quantum mechanics instead, Eq.~\eqref{eq:madelung1} and Eq.~\eqref{eq:madelung2}, the artificial viscosity term would be placed, together with the quantum force, in the momentum equation. Thus, Eq.~\eqref{eq:madelung2} would read:
\begin{align}
\frac{\partial v}{\partial t} + (\vec{v}\cdot \nabla)\vec{v} = -  \nabla \left ( V + \frac{g}{m^2}\rho + Q - \frac{\epsilon}{2m} \frac{ \nabla \left ( \rho\vec{v} \right )}{\rho} \right ).
\end{align}
The new viscous force term:
\begin{align*}
F_{\rm{viscous}} = -\frac{\epsilon}{2m} \nabla \left( \frac{ \nabla \left ( \rho\vec{v} \right )}{\rho} \right ),
\end{align*}
helps preventing high-frequency waves to build up in time. With the addition of such a viscosity term, we are able to evolve without any issues the wave-function over hundreds of oscillation periods in our tests and, at the same time, preserving mass, energy and agreement with analytical solutions. There is no unique prescription for solving these issues we encountered and, in general, it is possible to design more elaborate artificial viscosity terms. 

In an AMR context, when choosing the value of the dumping constant $\epsilon$, one should keep in mind that spurious high-frequency oscillations appear on scales of the local grid resolution, which is not fixed, but it changes according to designed refinement criteria. Thus, the strength of the dumping term should be decided such that it does not over-suppress the wave-function in high-resolution regions, but rather slightly under-suppress spurious oscillations in low-resolution regions. Such a limitation of the artificial viscosity term we implemented motivates further investigations to develop a more accurate scheme for dumping high-frequency spurious oscillations.
Empirically, we find that a value in the range $0.2 < \epsilon < 1$ ensures stability over a long time in all the test cases we present (except the soliton test case, were we set $\epsilon = 0$ and we do not use any artificial viscosity), by preventing the growth of spurious oscillations in the solution of the non-linear Sch\"odinger equation.

\subsection{Code units} 
We adopt the set of \textquotesingle super-comoving coordinates\textquotesingle ~introduced in \citet{1998MNRAS.297..467M}, and already used in \texttt{RAMSES}. Thus, the following change of variables is performed:
\begin{align}
\begin{gathered}
\tilde{x} ~= ~ \frac{x}{a L},~~~d\tilde{\tau} ~ = ~ \frac{H_0{\rm d}t}{a^2},~~~\tilde{\psi} ~ = ~ \frac{\psi}{\overline{\psi}}, \\~~~\tilde{V} ~ = ~ \frac{V a^2}{(H_0L)^2},~~~\tilde{m} ~ = ~ \frac{mH_0L^2}{\hbar},~~~\tilde{g} = \frac{a^2g}{H_0\hbar}|\overline{\psi}|^2,
\end{gathered}
\end{align}
where $H_0$ is the Hubble constant, $L$ is the box size, and $\overline{\psi}$ is chosen to ensure that $\int |\tilde{\psi}|^2 ~ {\rm d}^{\rm dim}\tilde{x} = 1$. As a consequence, the resulting non-linear Schr\"odinger equation reads:
\begin{align}
i\frac{d\tilde{\psi}}{d\tilde{\tau}} + \frac{1}{2\tilde{m}}\tilde{\nabla}^2\tilde{\psi} - \tilde{m}\tilde{V} \tilde{\psi} - \tilde{g}|\tilde{\psi}|^2\tilde{\psi} = 0.
\end{align}
This set of coordinates was specifically designed for cosmological applications. However, it can be used for any application by setting the scale factor $a$ to unity and replacing $H_0$ by a general inverse time scale $T^{-1}$. In the remainder of the paper, all equations are in these code units.

For the particular case of axion dark matter in a cosmological setting (see Eq.~\eqref{eq:axioncosmo}) we have $\tilde{g} = 0$, $\overline{\psi} \propto a^{-3/2}$, and the potential is determined via the Poisson equation:
\begin{align}
\tilde{\nabla}^2 \tilde{V} &= \frac{3}{2}a \Big [\Omega_{\rm axions}(|\tilde{\psi}|^2-1) + \Omega_{\rm CDM}(\tilde{\rho}_{\rm CDM}-1) \nonumber \\
&+ \Omega_{\rm baryons}(\tilde{\rho}_{\rm baryons}-1) + \ldots \Big ] ~ ,
\end{align}
where $\Omega_{i}$ is the fraction of the energy budget of our Universe that is in matter component $i$ (axions, baryons, CDM, etc.) and the mean value of $\tilde{\rho}_i$ over the box is set to unity.

\section{Tests of the code}\label{sec:tests}

In this section we present the numerical experiments we performed in order to test the main features of \codenamex. When testing the accuracy of our numerical schemes, we rely on three main tests: conservation of mass, energy, and reproduction of analytical solutions. Given the total mass $M\left( t \right)$ in the simulation box and the total energy $E\left( t \right)$, the corresponding errors are respectively defined as:
\begin{align*}
\epsilon_{\rm{mass}} = \left | \frac{M\left( t \right)-M\left( 0 \right)}{M\left( 0 \right)} \right | ~~~ \text{and} ~~~ \epsilon_{\rm{energy}} = \left | \frac{E\left( t \right)-E\left( 0 \right)}{E\left( 0 \right)} \right | ~ .
\end{align*}
Instead, by denoting the analytical solution as $\psi_{\rm{a}}\left(x,t\right)$ and the numerical solution as $\psi_{\rm{n}}\left(x,t\right)$, at a given time, the error with respect to the analytical solution is computed according to the following formula:
\begin{align*}
\epsilon_{\rm{solution}} =  \frac{\left<(\psi_{\rm{a}}\left(x,t\right)-\psi_{\rm{n}}\left(x,t\right))^2\right>_x}{\left<\psi^2_{\rm{a}}\left(x,t\right)\right>_x} ~ .
\end{align*}
where $\left<~~~\right>_x$ denotes the mean over the box. For the tests we present here, we always assume that each cell in the AMR grid has the same size in all dimensions: $\Delta x = \Delta y = \Delta z$. Furthermore, in all the tests, we only employ the refinement criterion based on density, as we found no need to use the L\"ohner error estimator in order to achieve an accurate and stable solution over time.

\paragraph{Accuracy and performances.}

While running the tests we discuss in this section, we measured the overall accuracy and performances of the numerical algorithms implemented in \codenamex.

\begin{figure}
\centering
\includegraphics[width=\columnwidth]{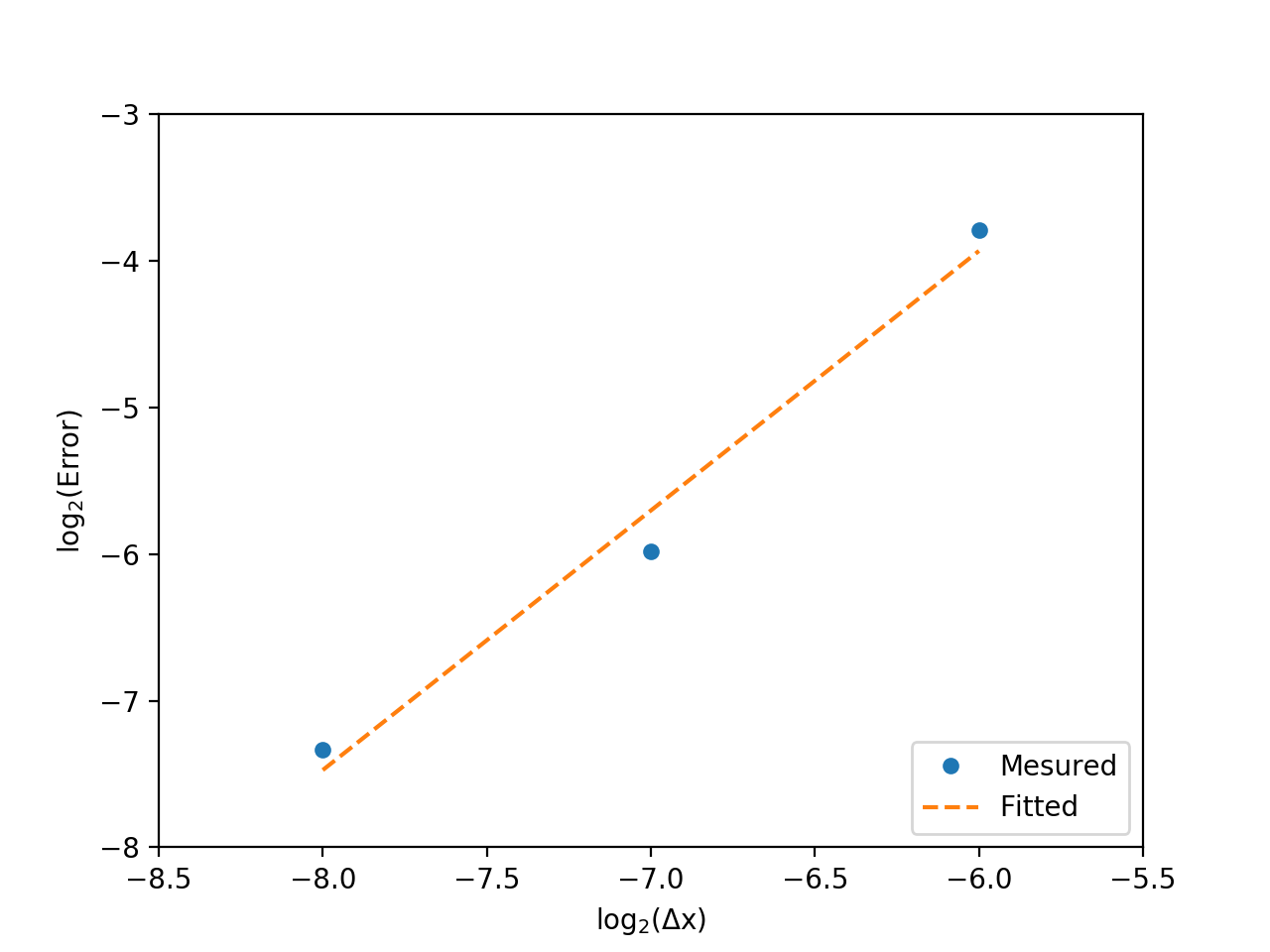}
\caption{
Overall accuracy of the numerical scheme implemented to solve the Schr\"odinger-Poisson system. Blue points represent errors with respect to the analytical solution for the soliton test, computed at a given time for $\Delta x = 2^{-8}$, $2^{-7}$, $2^{-6}$. The orange dashed line corresponds to the polynomial fit.}
\label{fig:accuracy}
\end{figure}

The accuracy of a numerical method is often measured by comparing the analytical solution with the numerical solution. 
For this purpose, we compute the error with respect to the analytical solution as given above. The overall accuracy of \codename is obtained by measuring how the global error, computed at a given time, scales with the grid spacing. In Fig.~\ref{fig:accuracy}, we show the sample of errors we obtained for one of the numerical experiments we performed, the soliton test. The data points are fitted by the formula:
\begin{align*}
\log_2 \left (y \right ) = \alpha \log_2\left( x \right) + \beta ~ ,
\end{align*}
where $x$ and $y$ represents the grid size and the error with respect to the analytical solution, respectively. The parameters of the fit $\alpha$ and $\beta$ are determined by means of the Linear Least Square (LLS) method and, in this case, we obtain $\alpha \sim 1.8$ and $\beta \sim 6.7$.
While $\beta$ is not relevant when measuring the convergence of a numerical scheme, the parameter $\alpha$ represents the slope of the fitting polynomial and it corresponds to the global accuracy of the solver. Thus, for \codenamex, we achieve an overall second-order accuracy.

We measured general performances of \codename on a small cluster. Every node of the cluster is equipped with two Intel E5-2670 processors, with 8 cores each, a clock frequency of $f_{\rm CPU} = 2.6$ GHz and a total memory of $128$ GB. Nodes within the cluster are interconnected through a Mellanox 56 Gb FDR Inifniband. In application with gravity enabled, \codename can evolve the solution of the non-linear Schr\"odinger equation by updating a single cell in the AMR hierarchy in $\sim11$ $\mu$s. This means that, our code is able to reach a value of $\sim 10^{5}$ cell updated per second. 

\paragraph{Conservation of mass.}
The non-linear Schr\"odinger equation has the conserved quantity:
\begin{align*}
M = \int |\psi|^2 ~ {\rm d}^{\rm dim}x,
\end{align*}
which in the Mandelung formulation is just the total mass of the fluid.
Mass and energy are not manifestly conserved by the main solver, therefore monitoring them is a useful test.

\begin{figure}
\centering
\includegraphics[width=\columnwidth]{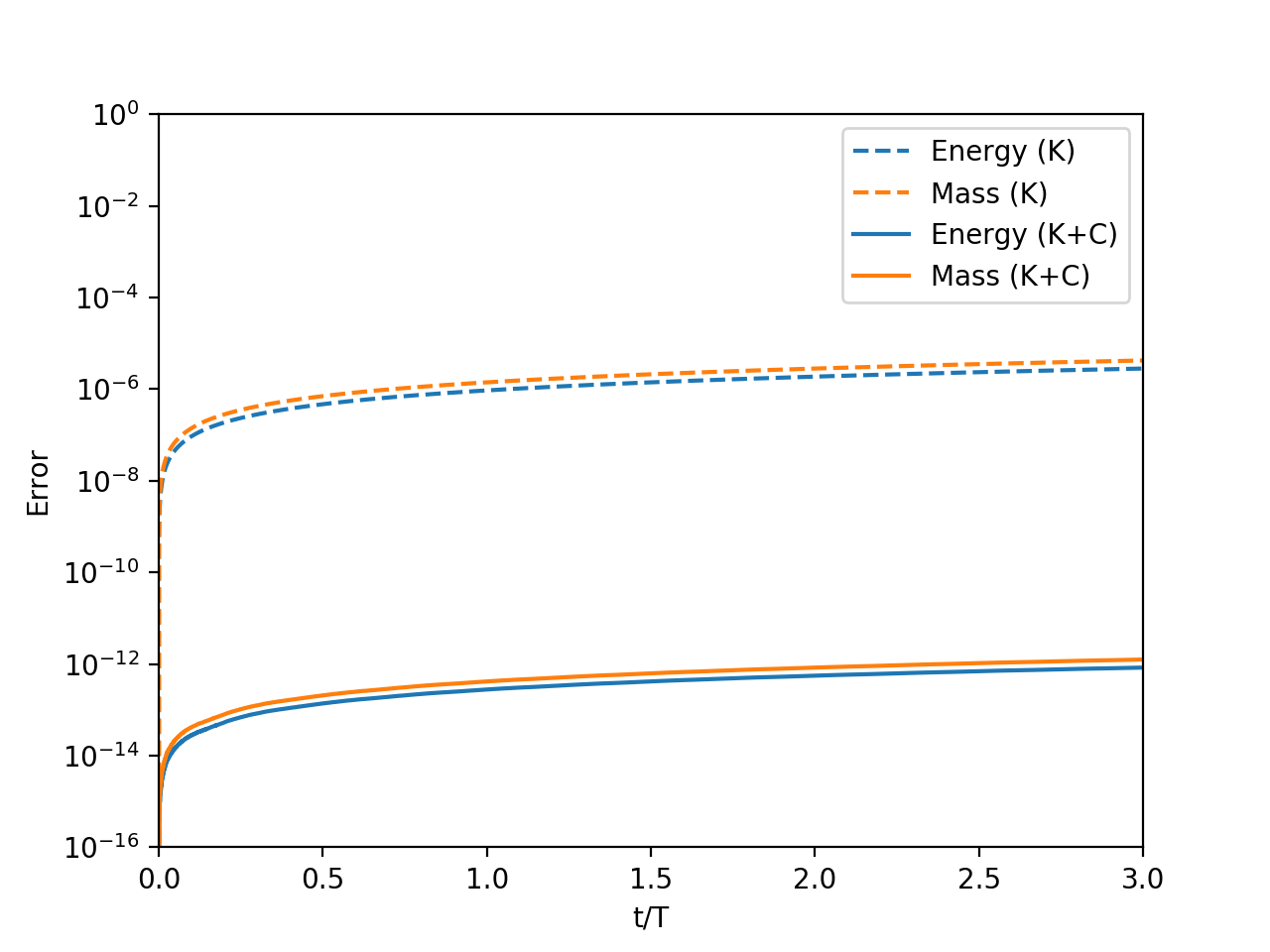}
\caption{
Errors in conservation of mass and energy when solving the Schr\"odinger equation only (K) and when solving together the Schr\"odinger and the continuity equations (K+C). Solving the continuity equation together with the Schr\"odinger equation improves the conservation properties of the algorithm by $\sim6$ orders of magnitude.}
\label{fig:mass}
\end{figure}

In simulations with no refinements and without enforcing mass conservation, we typically find the error on the conservation of mass of the order of (the prefactor is here for $\Delta x = 2^{-6}$):
\begin{align*}
\frac{\Delta M}{M} \sim 10^{-6}\left(\frac{t}{T} \right),
\end{align*}
where $T$ is the oscillation period. When we allow for refinements, the situation is typically worse and it is not good enough for cosmological simulations. However, by solving the continuity equation on top of the Schr\"odinger equation, we observe an improvement on the conservation of mass up to:
\begin{align*}
\frac{\Delta M}{M} \sim 10^{-13}\left(\frac{t}{T}\right).  
\end{align*}
This does not change when we allow refinements and, even though the error grows linearly in time, it is good enough in order to perform cosmological simulations. This is shown in Fig.~(\ref{fig:mass}), where we perform the sine wave test on the domain grid only, with a resolution of $N_{\rm cell} = 2^6$ cells, corresponding to $\ell=6$ and $\Delta x = 2^{-6}$, in one dimension. Details regarding the sine wave test are described in Section~\ref{sec:sinewave}.

Please, refer to Section~\ref{sec:numerical} for an explanation of the temporal evolution of the error in the conservation of mass.

\paragraph{Conservation of energy.}
Since we enforce mass conservation by solving the continuity equation, energy conservation is a better accuracy test for our code. 
By defining kinetic and potential energy as:
\begin{align}
K &\equiv \frac{1}{2m^2}\int |\nabla\psi|^2 ~{\rm d}^{\rm dim}x, \\
W &\equiv \frac{1}{2} \int V_{\rm eff}|\psi|^2 ~ {\rm d}^{\rm dim}x,
\end{align}
the temporal change in the total energy of the system is expressed by:
\begin{align}
\frac{d}{dt}(K+W) = \frac{1}{2}\int \frac{\partial V_{\rm eff}}{\partial t}|\psi|^2 ~ {\rm d}^{\rm dim}x.
\end{align}
As we can see, in the case where the effective potential has no explicit time-derivatives, the energy $E = K + W$ is conserved under the evolution.

In a cosmological setting, the potential $V$ depends on time via the scale-factor and this leads to a Lazyer-Irvine equation \citep{2017PhRvD..96l3532K}: 
\begin{align}\label{eq:LI}
\frac{d}{dt}(K+W) - HW = 0,    
\end{align}
which can be monitored by integrating it up while performing the simulation. 

\paragraph{Comparison to analytical solutions.}
The most stringent test we can perform is to directly compare the numerical solution with an analytical solution. However, the discretised version of the governing equations is a different problem than the theoretical continuous limit and, thus, it admits a different solution. Usually, the main difference between the solutions of the continuous and the discretised non-linear Schr\"odinger equations leads to the wave-function evolving with slightly different temporal phases. For this reason, we estimate the phase difference between the two solutions and, starting from the analytical solution of the continuous non-linear Schr\"odinger equation, we compute the theoretical solution of the discretised equation. Then, we compare the numerical solution with the proper solution of the discretised non-linear Schr\"odinger equation.
This represents a way of comparing the numerical solution with the one of the physical problem we are actually modelling. In the limit of $\Delta x \to 0$, the solutions of the continuous and the discretised non-linear Schr\"odinger equations converge to the same solution.
In this way, we show that the numerical solution is in excellent agreement with the theoretical solution of the discretised problem, even in the case when the resolution is not extremely high, where computation time would significantly increase. 

\hfill \newline \hfill
\hfill \newline \hfill
In the plots below we show, for different choices of initial conditions and potential, the error on the conservation of mass, the error on the conservation of energy, and the error with respect to the analytical solution.

\subsection{Sine wave}\label{sec:sinewave}
\begin{figure}
\centering
\includegraphics[width=\columnwidth]{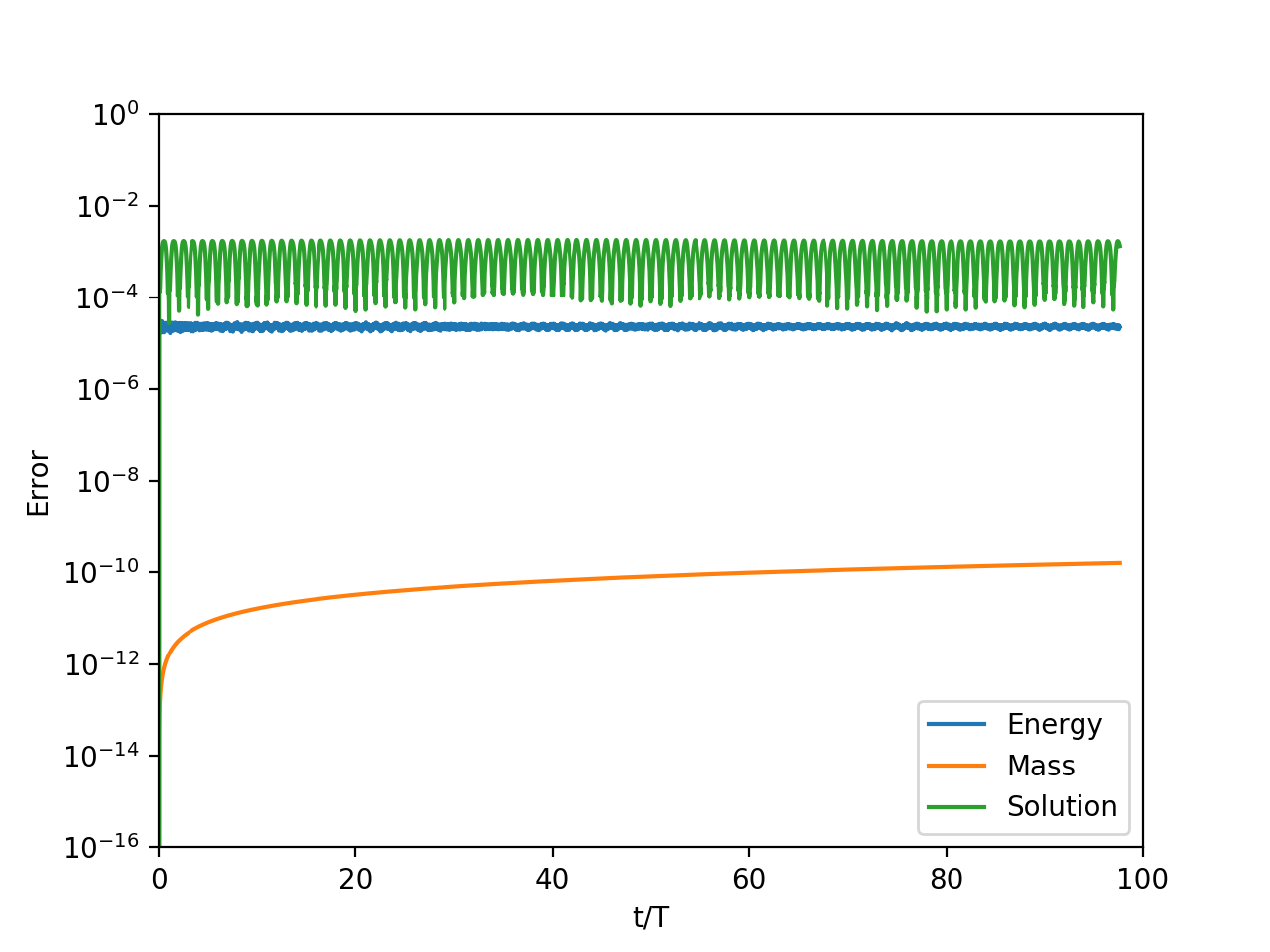}
\caption{Evolution of the three errors as a function of time for the cosine wave test. While the error on the conservation of mass (orange) evolves accordingly to $\Delta M / M \sim 10^{-13}(t/T)$, the error on the conservation of energy (blue) stays constant. Furthermore, the error with respect to the analytical solution (green) does not evolve over time.
}
\label{fig:coswavetest}
\end{figure}

The sine wave test evolves a static one dimensional density profile, where the initial wave-function is set as:
\begin{align}
\psi(x,0) = \sin(2\pi n x).
\end{align}
It evolves in a constant potential, which is defined as:
\begin{align}
V = 1 - \frac{2\pi^2 n^2}{m^2},
\end{align}
where $m$ is the mass carried by the wave-function and the period of oscillation is given by:
\begin{align}
T = \frac{2\pi}{m}.
\end{align}
The full analytical solution reads:
\begin{align}\label{eq:an_sol_sine}
\psi(x,t) = e^{-i\frac{2\pi t}{T}} \sin(2\pi n x).
\end{align}
This numerical experiment is designed to test the creation of ghost cells when computing fine levels boundary conditions. Here we only refine according to the mass criterion, Eq.~\eqref{eq:masscrit} and, since the density profile does not evolve in time, there is no dynamical creation or destruction of grids: once the refinement map is computed at the beginning, it does not change. We evolve the solution of the Schr\"odinger equation over $100$ periods of oscillation. 
It is possible to show that the solution of the discretised equation - the one we are solving - is the same as Eq.~\eqref{eq:an_sol_sine}, up to second order in space, but with a slightly different period of oscillation. Therefore, to factor out the dependence of the period with resolution (which we test seperately), we correct the analytical solution by replacing $T$ with $T_{\rm discrete}$, where:
\begin{align}\label{eq:periodsinwave}
\frac{T_{\rm discrete}}{T} &= \frac{1}{1 + T^2 n^2\left(\frac{1-\cos(2\pi n\Delta x)}{(2\pi n\Delta x)^2} - \frac{1}{2}\right)} \nonumber\\
&\approx \frac{1}{1-\frac{\pi^2T^2n^2}{12}\Delta x^2}.
\end{align}
This test was performed for ${\rm dim} = 1$. The non-linear Schr\"odinger equation is solved together with the continuity equation, in order to enforce conservation of mass. The domain grid resolution is $N_{\rm cell} = 64$, corresponding to $\ell=6$ and $\Delta x = 2^{-6}$, and the maximum refinement level is set to $\ell_{\rm max} = 8$. When boundary conditions for fine levels are needed, phase and density are interpolated in ghost cells by means of fourth-order conservative interpolation. We used artificial viscosity with $\epsilon = 0.2$. The results from this test are shown in Fig.~(\ref{fig:coswavetest}). 

We have also performed a similar test using a quadratic potential leading to a Gaussian profile:
\begin{align*}
\psi \propto \exp\left(-\frac{x^2}{\sigma^2}\right),
\end{align*}
with very similar results.

\subsection{Travelling wave}
\begin{figure}
\centering
\includegraphics[width=\columnwidth]{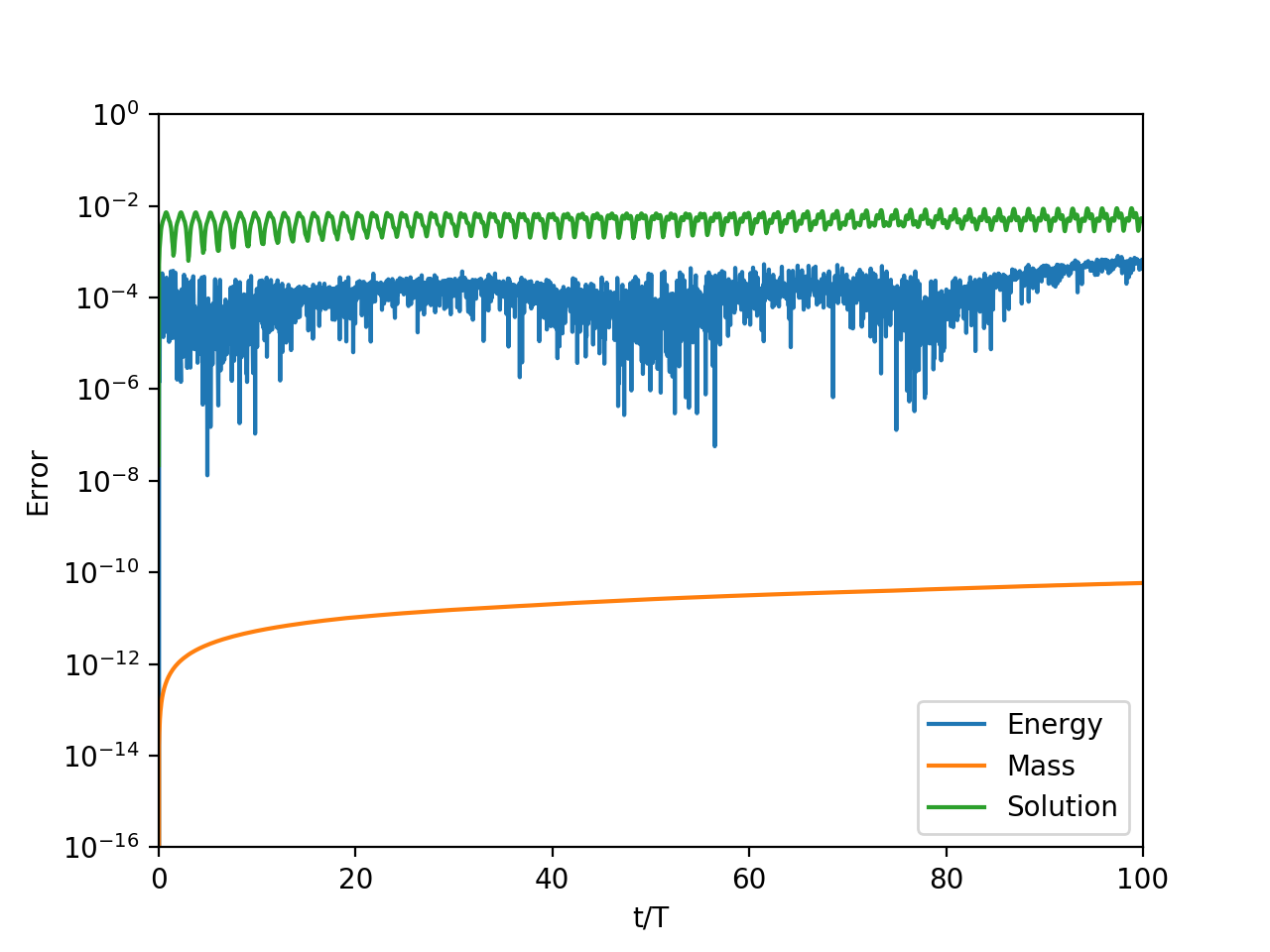}
\caption{Evolution of the three errors as a function of time for the travelling wave test. While the error on the conservation of mass (orange) evolves accordingly to $\Delta M / M \sim 10^{-13}(t/T)$, the error on the conservation of energy (blue) stays roughly constant. Furthermore, the error with respect to the analytical solution (green) does not evolve over time.
}
\label{fig:travwavetest}
\end{figure}

This test simulates a one dimensional wave-packet travelling through a periodic box. Here, we test dynamical creation and destruction of grids, since the AMR hierarchy follows the density profile moving towards the direction of the wave. In this case, we have no potential and the initial conditions are defined as:
\begin{align}
\psi(x,0) = \frac{1}{\sqrt{2}}\left[e^{i k_1x} + e^{i k_2x}\right],
\end{align}
where $k_1 = 2\pi n_1$, $k_2 = 2\pi n_2$ with $n_1\not= n_2\in\mathbb{N}$. The oscillation frequency of a single mode is:
\begin{align*}\omega(k) = \frac{k^2}{2m},
\end{align*}
and the analytical solution of the Schr\"odinger equation reads:
\begin{align}
\psi(x,t) = \frac{1}{\sqrt{2}}\left[e^{i(k_1x - \omega(k_1) t)} + e^{i(k_2x - \omega(k_2) t)}\right].
\end{align}
As a consequence, the density is given by:
\begin{align}
|\psi(x,t)|^2 = 1 + \cos \left(2\pi x(n_2-n_1) + \frac{2\pi t}{T}\right),
\end{align}
where the oscillation period is defined as:
\begin{align}
T = \frac{m}{\pi(n_1^2-n_2^2)}.
\end{align}
The wave-function is evolved in time over $100$ oscillation periods. 
The non-linear Schr\"odinger equation is solved together with the continuity equation, in order to enforce conservation of mass. 
Also in this case, the coarse-fine data interpolation is performed by means of fourth-order conservative interpolation. However, while density and phase are interpolated in ghost cells, new refinements are made by interpolating real and imaginary parts of the wave-function. The domain grid has the same resolution as in the previous test and refinements are allowed up to $\ell_{\rm max} = 8$. We used artificial viscosity with $\epsilon = 0.2$. The results from this test are shown in Fig.~(\ref{fig:travwavetest}).

\subsection{Soliton}
In a cosmological context, \codename can be used to simulate the structure formation process with fuzzy dark matter. In this case, the density profiles of the dark matter halos differs from the case of the standard CDM. 
We can find a stationary solution which can be tested by taking:
\begin{align*}
\psi(x,t) = e^{-i\frac{2\pi t}{T}}\chi(x),
\end{align*}
and solving the resulting ODE for $\chi(x)$. A numerical fit to the density profile of a soliton in three dimensions was first suggested in \citet{2014NatPh..10..496S} and then in \citet{2015MNRAS.451.2479M}. Despite it is an approximated solution, it is useful when coupling the Poisson equation to the non-linear Schr\"odinger equation. In this works, the density profile of the soliton was found to be on the form:
\begin{align}
\rho \propto \frac{1}{\left[1 + (r/r_{\rm core})^2\right]^8},
\end{align}
where $r_{\rm core}$ can be chosen as a free parameter, see Appendix~\ref{sect:app_soliton} for more details.

\begin{figure}
\centering
\includegraphics[width=\columnwidth]{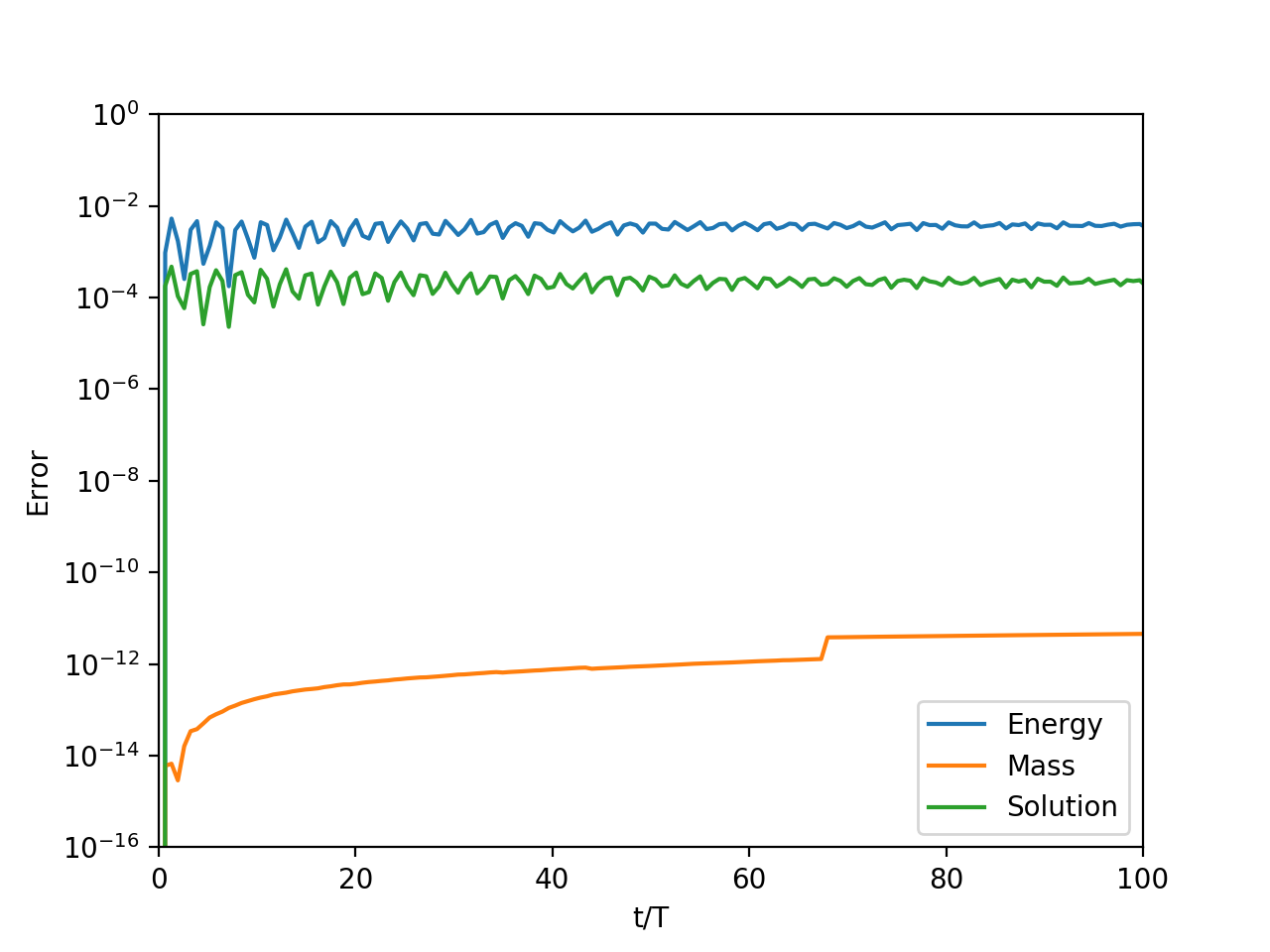}
\caption{Evolution of the three errors as a function of time, for a self-gravitating soliton. While the error on the conservation of mass (orange) evolves accordingly to $\Delta M / M \sim 10^{-13}(t/T)$, the error on the conservation of energy (blue) stays constant. Furthermore, the error with respect to the analytical solution (green) does not evolve over time.
}
\label{fig:solitontest}
\end{figure}

We set this density profile analytically and evolve the system. The density profile remains approximately stationary while the wave-function oscillates as:
\begin{align*}
\psi(x,t) \propto e^{-i\frac{2\pi t}{T}}.
\end{align*}
This test was performed for $\rm{dim} = 3$ and the non-linear Schr\"odinger equation is solved together with the continuity equation. The domain grid contains $N_{\rm cell}= 64^3$ cells, corresponding to $\ell=6$ and $\Delta x = 2^{-6}$, and the maximum refinement level allowed is $\ell = 8$. In this case, both refinement and ghost cells are made by fourth-order conservative interpolation on density and phase. 
The artificial viscosity term is set to $\epsilon=0$. We found that self-gravity is able to stabilise the wave-function against spurious numerical oscillations. We tested the same physical case with different values of artificial viscosity,  but we did not find any improvement in terms of accuracy. In the other tests, where we set $\epsilon=0.2$, the artificial viscosity term was introduced just in order to ensure stability and accuracy of the solution to the non-linear Schr\"odinger equation over a very large number of oscillation periods.
The results for this test are shown in Fig.~(\ref{fig:solitontest}).

\section{Cosmological applications}\label{sec:cosmo}

\begin{figure}[t]
\centering
\includegraphics[width=\columnwidth]{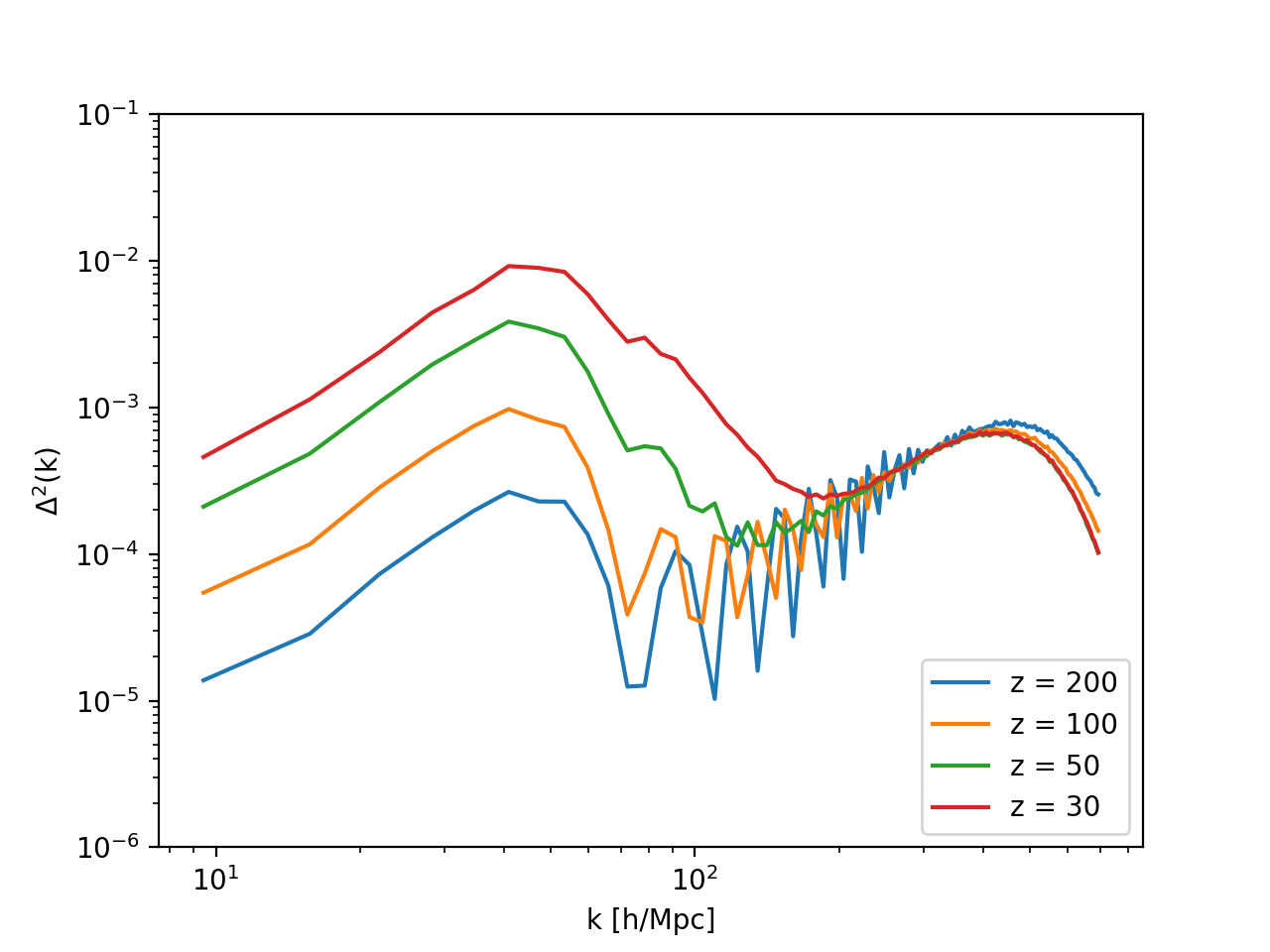}
\caption{
Evolution of the dimensionless power spectrum $\Delta^2(k)$ with redshift.
}\label{Fig:pofk}
\end{figure}

\begin{figure}[t]
\centering
\includegraphics[width=\columnwidth]{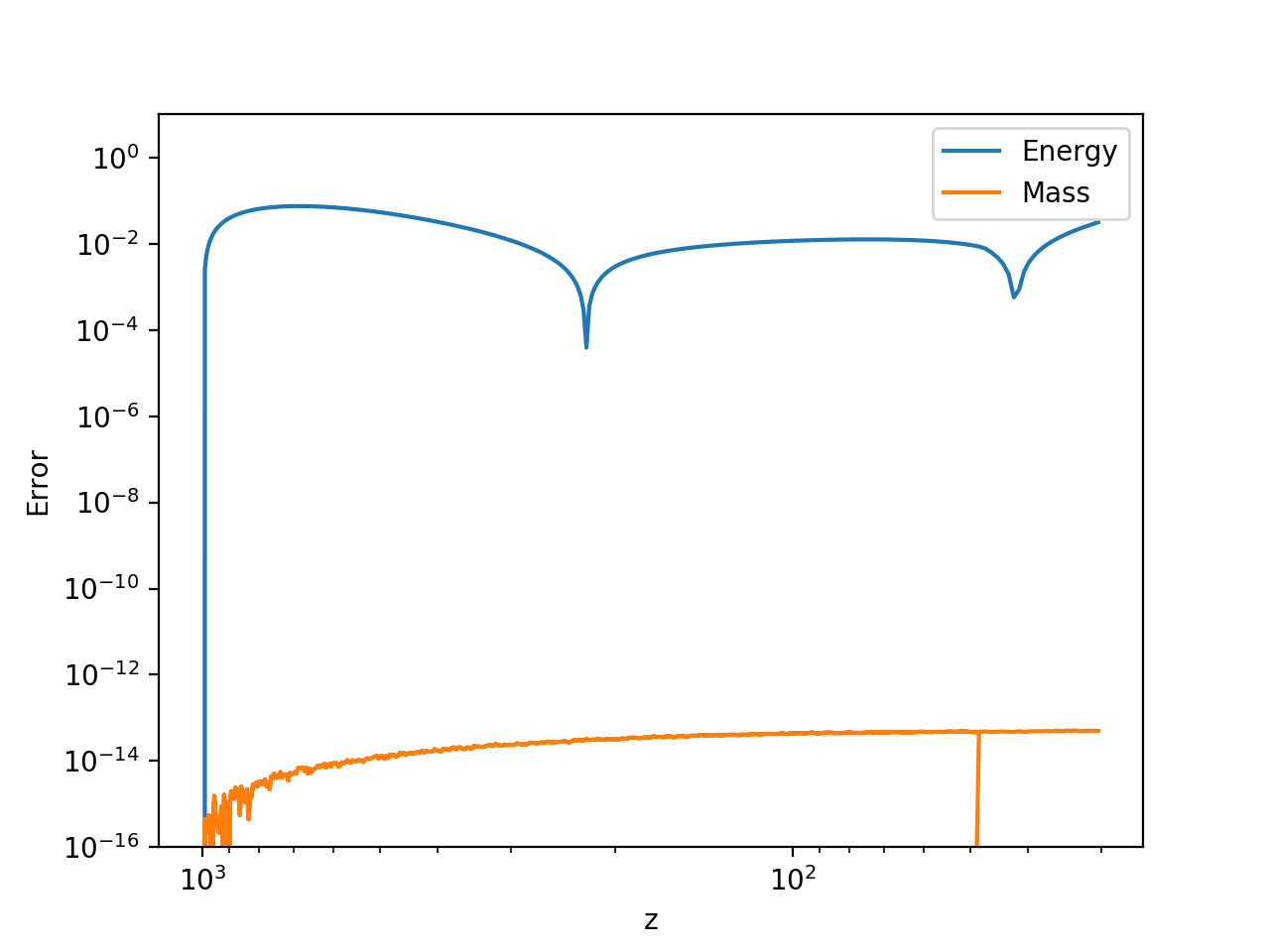}
\caption{
Evolution of the errors on conservation of mass (orange) and energy (blue), as a function of redshift.
}\label{Fig:errs}
\end{figure}

\begin{figure*}[!htb]
\centering
\includegraphics[width=\linewidth]{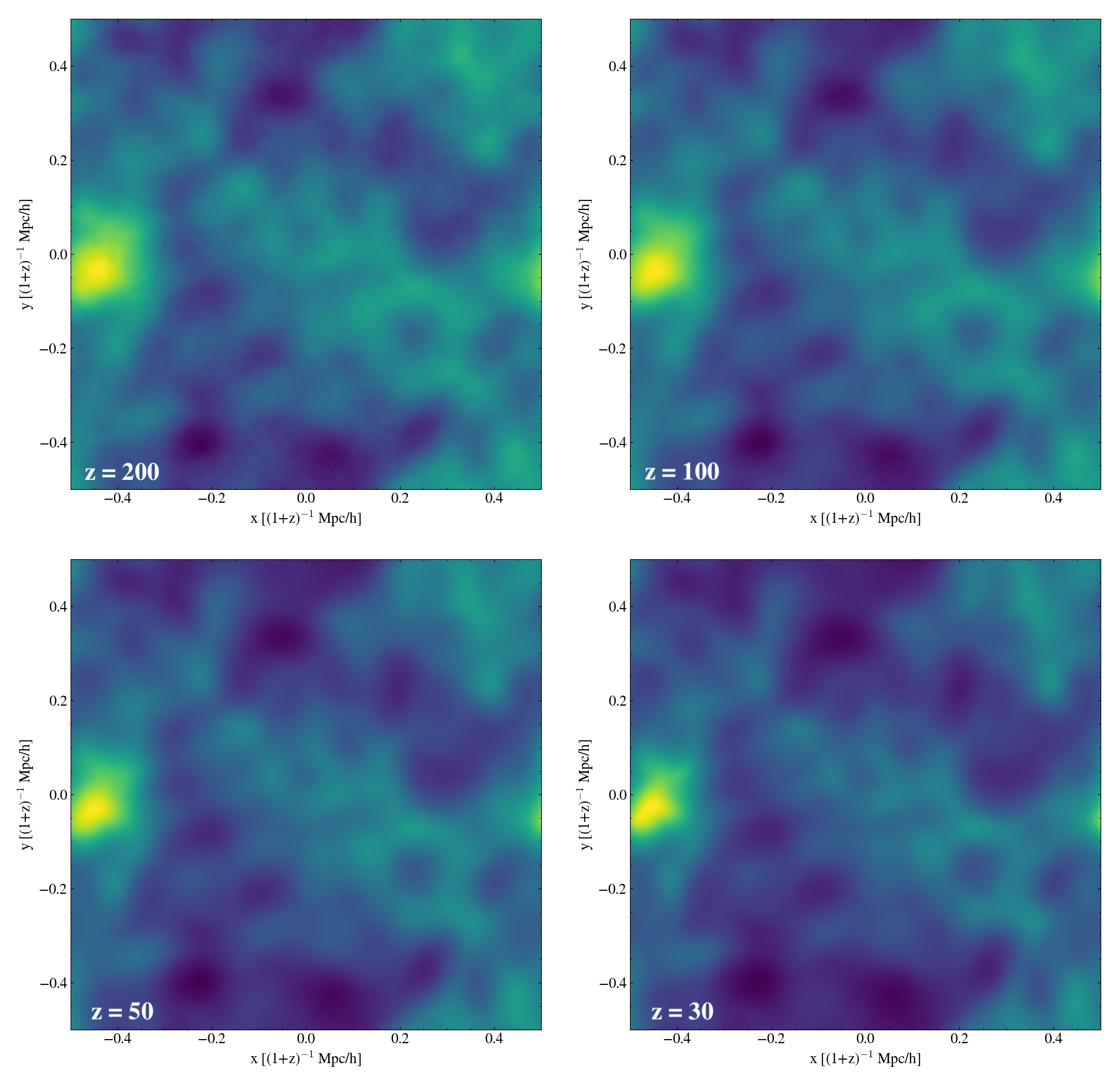}
\caption{
Projection along the $z$ axes of the dark matter density field, normalised by the critical density of the Universe. The box is $1~\text{Mpc}/h$ in comoving units and it represents the entire simulation box.
}\label{Fig:ax_dens}
\end{figure*}
\begin{figure*}[!htb]
\centering
\includegraphics[width=\linewidth]{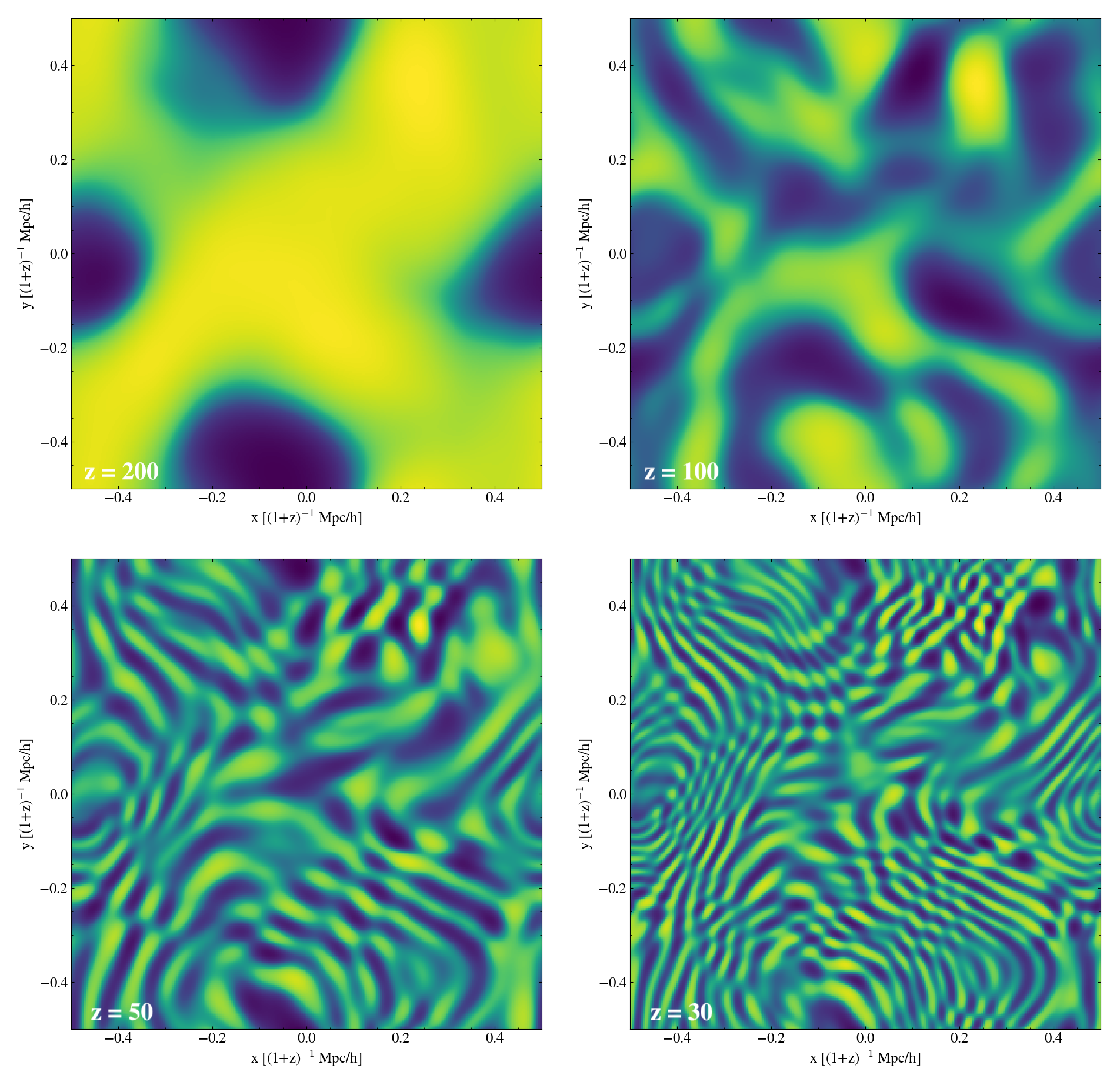}
\caption{
Projection along the $z$ axes of the real part of the dark matter field. The box is $1~\text{Mpc}/h$ in comoving units and it represents the entire simulation box.
}\label{Fig:ax_real}
\end{figure*}
\begin{figure*}[!htb]
\centering
\includegraphics[width=\linewidth]{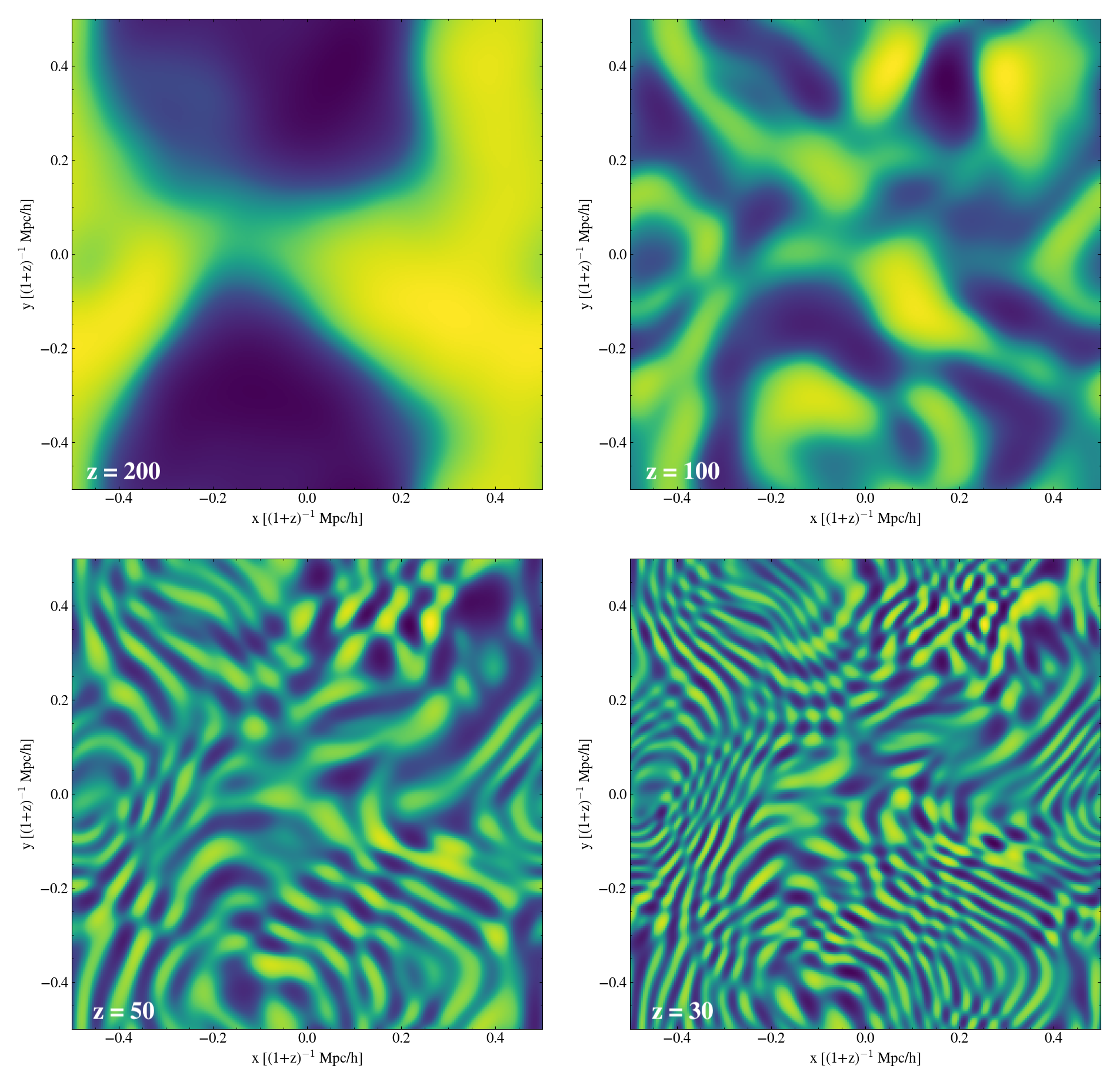}
\caption{
Projection along the $z$ axes of the imaginary part of the dark matter field. The box is $1~\text{Mpc}/h$ in comoving units and it represents the entire simulation box.
}\label{Fig:ax_imag}
\end{figure*}

The \codename code was originally developed in order to perform numerical simulations of structure formation with fuzzy dark matter. To demonstrate the capabilities of our code, in this section we present a test involving a full 3D cosmological setup.

The simulation box models a $B_0 = 1~\rm{Mpc}/h$ portion of the Universe expanding over time, according to the Einstein-de Sitter model. We take $\Omega_{\rm{\Lambda}} = 0.7$ and $\Omega_{\rm{M}} = 0.3$, and the Hubble constant is set to $H_0 = 100 ~h~\rm{km}~\rm{s}^{-1}~\rm{Mpc}^{-1}$, where $h=0.67$. The base resolution, which defines the domain level, is set to $N_{\rm cell} = 256^3$ and up to one level of refinement is allowed. This means that, in this test, we achieve a maximum resolution of $\Delta x \sim 4~h^{-1}\rm{kpc}$. The mass of the boson is set to $m = 10^{-21}~\rm{eV}$ and conservation of mass is enforced by solving the continuity equation on top of the non-linear Schr\"odinger equation.
In order to better appreciate the differences between CDM and fuzzy dark matter, the initial conditions are computed for the case of CDM. 
The initial density and velocity fields are computed by using the Zel'dovich approximation. Then, following \citet{2017PhRvD..96l3532K}, we convert them into an initial wave-function.

The evolution starts at redshift $z=1000$ and we run the simulation as long as we are able to resolve the quantum force with at least two cells, meaning until redshift $z=30$. We did not run this simple test further because, due to lack of resolution, we cannot resolve the inner part of collapsed objects. This will be the subject of an upcoming paper.

In Fig.~\ref{Fig:ax_dens}, Fig.~\ref{Fig:ax_real} and Fig.~\ref{Fig:ax_imag} we show the dark matter density field, the real and the imaginary parts of the wave-function for a selection of redshifts, $z = 200, 100, 50, 30$. As the field clusters under the effect of gravity, the wave-function develops the wave patterns which are characteristic of this class of models.

By taking the density contrast $\delta \left( \mathbf{x} \right)$ inside the simulation box, we expand it in Fourier modes as follows:
\begin{align}
\delta\left( \mathbf{x} \right) = \int d^3 k ~ \delta \left( \mathbf{k} \right) \exp\left( -i \mathbf{k} \cdot \mathbf{x} \right).
\end{align}
The matter power spectrum is defined by means of the autocorrelation function, which can be expressed as:
\begin{align}
\braket{\delta \left( \mathbf{x} \right)\delta \left( \mathbf{x} \right)} &= \int_0^{\infty} \frac{dk}{k} \frac{k^3 \left| \delta\left( k \right)\right|^2}{2\pi^2} \\
												  &= \int_0^{\infty} \frac{dk}{k} \frac{k^3 P\left( k \right)}{2\pi^2}.
\end{align}
In Fig.~\ref{Fig:pofk}, we plot the dimensionless power spectrum, defined as:
\begin{align}
\Delta^{2}	\left( k \right) = \frac{k^3 P\left( k \right)}{2 \pi^2},
\end{align} 
for $z = 200, 100, 50, 30$. The results here show the same quantitative behaviour as seen in \citet[see Fig. 2]{2009ApJ...697..850W}, that performed the same kind of simulation as we do here.  As the field gravitationally collapses, the quantum pressure leaves immediately its imprints on $\Delta \left( k \right)$ by producing the characteristic suppression of power at small scales, above $k \sim 300~h^{-1}\text{Mpc}$. Given the mass of the boson, the suppression scale is expected to be around the redshift dependent Jeans wave-number, which can be defined as:
\begin{align} \label{eq:supp_scale2}
k_{J} &= \left( \frac{16 \pi G \rho_{a}}{1+z} m^2 \right)^{1/4} \nonumber \\
&= 66.5 \left( \frac{\Omega_{a}}{0.12 h^2} \right)^{1/4} \left( \frac{m}{10^{-22}~{\rm eV}} \right)^{1/2} \left( 1+z \right)^{-1/4} ~ \frac{h}{\text{Mpc}}~,
\end{align}
where $\Omega_a$ is the dimensionless density parameter of axions and $h$ is the dimensionless Hubble constant.
Between $k = 10~h^{-1}\text{Mpc}$ and $k = 100~h^{-1}\text{Mpc}$, the power spectrum describes modes in the density field still in the linear regime and, therefore, $\Delta^2 \left( k \right)$ evolves with redshift according to linear theory: 
\begin{align*}
\Delta^2 \propto \left( 1+z \right)^{-2},
\end{align*}
in a similar way to the CDM case.

In Fig.~\ref{Fig:errs}, the evolution of errors in conservation of mass and energy are plotted against redshift. The error in the conservation of mass slowly evolves in time, as described in Section \ref{sec:tests}. 
Furthermore, we track the evolution of the error in the conservation of energy, by integrating the Lazyer-Irvine equation along the simulation and checking at which level of accuracy Eq.~\eqref{eq:LI} is satisfied. As shown in the same figure, in the conservation of energy does not grow significantly, thus ensuring that no energy is numerically dissipated by the solver or the artificial viscosity term.

\section{Conclusions}\label{sec:conc}
In \codenamex, we implemented a set of numerical algorithms developed in order to solve the non-linear Schr\"odinger equation in an AMR framework. Eq.~\eqref{eq:GPE} can be used to describe the dynamics of a Bose-Einstein condensate, a system of identical bosons in the ground energy state, by means of a single-particle wave-function in a mean field approach. Here, the non-linearity arises from an effective potential, which can contain both a self-interaction term and a generic external potential. Bose-Einstein condensates find their application in several fields. As an example, alternative dark matter models involving Bose-Einstein condensates have been recently developed, such as ultra-light axion dark matter, fuzzy dark matter and superfluid dark matter. The Schr\"odinger equation is solved with a Taylor method, similar to the algorithm developed in \texttt{GAMER}. In order to improve the conservation properties of the numerical scheme, the continuity equation is solved on top of the non-linear Schr\"odinger equation and mass conservation is enforced by construction. 
Empirically, by running several tests, we found that our numerical method is second-order accurate. 

In order to test the main components of \codenamex, a test suite was designed. In particular, we tested the performances of the solver with and without solving the continuity equation on top of the non-linear Schr\"odinger equation, the creation of ghost cells when boundary conditions need to be specified for levels with fine resolution and the dynamical creation and destruction of grids during the process of mesh refinement. For this purpose, we tracked mass and energy conservation properties of our numerical schemes during the evolution of the system. The result is that both mass and energy are well conserved. While the latter remains roughly constant in all the different cases, the former evolves in time. Indeed, the error on the conservation of mass grows according to:
\begin{align*}
\frac{\Delta M}{M} \sim 10^{-6} \left( \frac{t}{T} \right),
\end{align*} 
for all our test-cases as we advance the solution in time. However, by solving the continuity equation on top of the non-linear Schr\"odinger equation, the error on the conservation of mass improves of several orders of magnitude, growing as: 
\begin{align*}
\frac{\Delta M}{M} \sim 10^{-13} \left( \frac{t}{T} \right),
\end{align*}
and it remains significantly small even for cosmological simulations. Furthermore, we compared the numerical solutions found by \codename with the analytical solutions of all the test cases. We show that the numerical solution tracks very well the analytical one over a long evolution time. In this case also, the error with respect to the analytical solution remains roughly constant over time. We showed that high-frequency spurious oscillations created at coarse-fine boundaries by interpolation schemes are efficiently dumped by an artificial viscosity term. However, the long term evolution of the single-particle wave-function still represents a challenge in case the artificial viscosity term is not included. We also run a small cosmological simulation where we show that \codename is able to capture the relevant features of models like fuzzy dark matter on cosmological scales.

Future work will aim at developing new high-order interpolation schemes which will not require the inclusion of an artificial viscosity term. Furthermore, we plan to compare the performance and accuracy of \codename with similar codes. In a following paper, we plan to exploit \codename to explore the non-linear regime of the structure formation with alternative dark matter models. In particular, we want to run a set of high-resolution cosmological simulation in order to verify and provide further predictions of fuzzy dark matter. 

The code will be soon publicly available through our GitHub repository\footnote{http://github.com/mattiamina}.

\begin{acknowledgements}
We thank the Research Council of Norway for their support. Computations were performed on resources provided by UNINETT Sigma2 -- the National Infrastructure for High Performance Computing and Data Storage in Norway. HAW was supported by the European Research Council through 646702 (CosTesGrav). We also thank the anonymous referee for several suggestions that helped improve the paper.
\end{acknowledgements}

\bibliography{references.bib}

\appendix
\section{Stability analysis}\label{sect:app_stability}
The stability condition of a generic PDE solver can be easily found by means of the Von Neumann stability analysis \citep{doi:10.1111/j.2153-3490.1950.tb00336.x}. For this purpose, the numerical error is decomposed in Fourier modes and a condition on time-step is computed in order to propagate each mode accurately.

We start by considering a generic mode in the Fourier decomposition of the numerical error:
\begin{align}
\epsilon_m = e^{at} e^{ik_m x}.
\end{align}
As mentioned in the previous sections, in \codename we use a second-order finite difference formula in order to approximate the laplacian of the wave-function. Thus, for a generic quantity, in one physical dimension we have:
\begin{align}
\nabla^2 f = \frac{f_{i+1}+f_{i-1}-2f_i}{\Delta x^2}.
\end{align}
As a consequence, the second derivative of the error can be written as:
\begin{align}
\nabla^2 \epsilon = -\frac{4}{(\Delta x)^2}\sin^2\left(\frac{k\Delta x}{2}\right)\epsilon,
\end{align}
and the amplification factor can be computed as:
\begin{align}
\xi &= \frac{\epsilon(t+ \Delta t)}{\epsilon(t)} \\
 &=e^{-imV \Delta t}\left[1 + i\beta - \frac{\beta^2}{2!} - \ldots + \frac{(i\beta)^n}{n!} \right],
\end{align}
where $\beta$ corresponds to:
\begin{align}
\beta &= -\frac{2 \hbar \Delta t}{m(\Delta x)^2}\sin^2\left(\frac{k\Delta x}{2}\right).
\end{align}
In order to avoid exponential growth, we require that $|\xi|<1$. Therefore, the stability condition reads:
\begin{align}
\left |\xi \right |^2 = \cos_n^2(\beta) + \sin_n^2(\beta) < 1,
\end{align}
where $\cos_n$ and $\sin_n$ denote to $n^{\rm th}$ order Taylor polynomials of $\cos(x)$ and $\sin(x)$, respectively. Furthermore, $n$ corresponds to the order of the Taylor expansion of the kinetic contribution to the time evolution operator, Eq.\eqref{eq:taylor_K}. 

In particular, we find that for $n<3$ the numerical scheme is unconditionally unstable. For $n=3$, instead, the stability condition is satisfied as long as:
\begin{align}
\left |\beta \right | < \sqrt{3} \implies  \Delta t < C_K\frac{\sqrt{3}}{2\hbar}m(\Delta x)^2.
\end{align}
The generalisation to $D$  can be done by replacing  $(\Delta x)^2\to (\Delta x)^2/D$ in the formula above.

We also require that the phase angle does not rotate more than $2\pi C_W$ within a time-step. Thus, for the kinetic term we require that:
\begin{align}
 \Delta t < C_{K}\cdot \frac{\pi m(\Delta x)^2}{\hbar}.
\end{align}
while for the potential term:
\begin{align}
\Delta t < C_W\cdot \frac{2\pi \hbar}{m|V_{\rm max}|}.
\end{align}
Combining the three conditions above, the optimal time-step is chosen as:
\begin{align}
 \Delta t < \min\left[C_{K} \cdot \frac{\sqrt{3}}{2\hbar} m\left(\Delta x\right)^2,~~C_W\cdot \frac{2\pi \hbar}{m|V_{\rm max}|}\right],
\end{align}
where we require the safety factors to be $C_W,C_K < 1$.

\section{Prolongation operators}\label{sect:app_prolong} 
The details related to the interpolation schemes we implement in \codename are given for the one dimensional case. In case of multidimensional interpolation, the same formulas derived in this section are applied sequentially in each direction.

\begin{figure*}
\centering
\includegraphics[width=2\columnwidth]{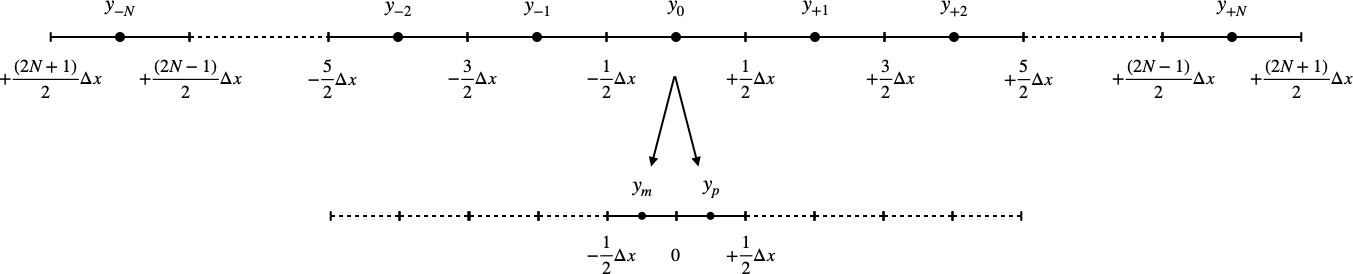}
\caption{Coarse-fine grid interpolation.}
\label{fig:interp}
\end{figure*}

For simplicity, in order to derive the interpolation formulas we use in \codenamex, we assume that there is an odd number $2N+1$ of coarse data points $\{ x_i, y_i \}$ with $-N \le i \le N$ and the interpolation is always done for children of the central cell, as shown in Fig.~(\ref{fig:interp}).

In this way, we can construct an interpolating polynomial of order $2N$ starting from its general definition:
\begin{align} \label{eq:poly4}
P(x) = \sum_{n=-N}^{N} a_n x^{n}.
\end{align}
In order to find an explicit expression for $P(x)$, the coefficients $\{a_n\}$ have to be computed and, depending on the constraints imposed on the interpolating polynomial, different schemes can be implemented. 

\subsection*{Lagrange polynomial interpolation}
For this interpolation scheme, instead of explicitly computing the coefficients $a_n$, the interpolating polynomial is expressed as a linear combination of Lagrange basis functions $l_j(x)$, with $j = -N, -N+1, \dots, N-1, N$, which are defined as:
\begin{align} \label{eq:lagrbasis}
l_j(x) = \prod_{\substack{m = -N \\ m \neq j}}^{N} \frac{x-x_m}{x_j-x_m}. 
\end{align}
Thus, coarse data points $y_i$ weight the Lagrange basis functions and the interpolating polynomial is constructed as follows:
\begin{align} \label{eq:lagrint}
P(x) = \sum_{n=-N}^{N} y_n ~ l_n(x).
\end{align}
In this way, the interpolating polynomial is forced to pass through the data points in the sample.
In order to find an explicit formula for a fourth-order Lagrange interpolating polynomial, we directly compute the Lagrange basis functions by means of Eq. \eqref{eq:lagrbasis}. Here, we express the differences between cell positions in Eq. \eqref{eq:lagrbasis} as a function of $\Delta x$. Then, the interpolating polynomial is computed at children cell positions, leading to the following interpolation formulas:
\begin{align}
y_m &= \frac{-45 ~y_{-2} + 420 ~y_{-1} + 1890 ~y_0 - 252 ~y_{+1} + 35 ~y_{+2}}{2048}, \\
y_p &= \frac{+35 ~y_{-2} - 252 ~y_{-1} +1890 ~y_0 + 420 ~y_{+1} - 45 ~y_{+2}}{2048}. 
\end{align}

\subsection*{Conservative polynomial interpolation}
In this case, constraints on the coefficients of the interpolating polynomial are set by imposing that the mean of the interpolated data on fine cells is equal to the data stored in the coarse cell.  

By using Eq. \eqref{eq:lagrint}, the coarse data points are redefined by means of their cell average:
\begin{align}
\tilde{y}_i = \frac{1}{\Delta x} \int_{\frac{2i-1}{2}\Delta x}^{\frac{2i+1}{2}\Delta x} P(x) ~ {\rm d}x,
\end{align}
and the corresponding linear system is solved in order to derive an explicit expression for the coefficients $\{ a_n \}$ of the interpolating polynomial. Thus, assuming that the parent cell is split into two children cells, the fine data is obtained by solving the following integrals:
\begin{align}
\tilde{y}_m &= \frac{1}{\Delta x} \int^{0}_{-\frac{\Delta x}{4}} P(x) ~ {\rm d}x, \\
\tilde{y}_p &=  \frac{1}{\Delta x} \int^{+\frac{\Delta x}{4}}_{0} P(x) ~ {\rm d}x,
\end{align}
where $\tilde{y}_m$ and $\tilde{y}_p$ denote the left and right childern cells respectively.

For a fourth-order polynomial, the solution of the linear system reads:
\begin{align}
a &= \frac{1}{(\Delta x)^4} \left [ \frac{ ~\tilde{y}_{-2} - 4~\tilde{y}_{-1} + 6~\tilde{y}_{0} - 4~\tilde{y}_{+1} + ~\tilde{y}_{+2} }{24} \right ], \\
b &= \frac{1}{(\Delta x)^3} \left [ \frac{ -~\tilde{y}_{-2} + 2~\tilde{y}_{-1} - 2~\tilde{y}_{+1} + ~\tilde{y}_{+2} }{12} \right ], \\
c &= \frac{1}{(\Delta x)^2} \left [ \frac{ -~\tilde{y}_{-2} + 12~\tilde{y}_{-1} - 22~\tilde{y}_{0} + 12~\tilde{y}_{+1} - ~\tilde{y}_{+2} }{ 16} \right ], \\
d &= \frac{1}{(\Delta x)} \left [ \frac{ 5~\tilde{y}_{-2} -34~\tilde{y}_{-1} + 34~\tilde{y}_{+1} - 5~\tilde{y}_{+2} }{48} \right ], \\
e &= \frac{ 9~\tilde{y}_{-2} - 116~\tilde{y}_{-1} + 2134~\tilde{y}_{0} - 116~\tilde{y}_{+1} + 9~\tilde{y}_{+2} }{1920}.
\end{align}
Thus, the corresponding interpolation formulas are:
\begin{align}
y_m &= ~\tilde{y}_0 - \frac{3 ~\tilde{y}_{-2} - 22 ~\tilde{y}_{-1} + 22 ~\tilde{y}_{+1} - 3 ~\tilde{y}_{+2}}{128}, \\
y_p &= ~\tilde{y}_0 + \frac{3 ~\tilde{y}_{-2} - 22 ~\tilde{y}_{-1} + 22 ~\tilde{y}_{+1} - 3 ~\tilde{y}_{+2}}{128}.
\end{align}
It is trivial to check the arithmetic average of $y$ over the children cells corresponds exactly to the value stored in the parent cell.

\section{Soliton solutions}\label{sect:app_soliton}
Self-gravitating bosonic fields can support stable and localised field configurations, where the density profile is static. Such configurations, called solitons, are ubiquitous in models of axion dark matter and exist for $\rm dim > 1$. Starting with the Schr\"odinger-Poisson system (in code units):
\begin{align}
&i\dfrac{\partial \psi}{\partial t} = -\dfrac{1}{2m}\nabla^2\psi + mV,  \\
&\nabla^2V = \kappa |\psi|^2,
\end{align}
we take the spherically symmetric ansatz: 
\begin{align} \label{eq:sph_ansaz}
\psi(r,t) = \chi(r) e^{-2\pi i\frac{t}{T}}.
\end{align}
The ODE describing the static configuration of the field can be derived by replacing Eq. \eqref{eq:sph_ansaz} into the Schr\"odinger-Poisson system:
\begin{align}
\nabla^2\left[\frac{\nabla^2 \chi}{\chi}\right] = 2 \kappa m^2 \chi^2,
\end{align}
where the laplacian is now expressed in polar coordinates:
\begin{align*}
\nabla^2 = \frac{d^2}{dr^2} + \frac{2}{r}\frac{d}{dr},
\end{align*}
and $\kappa$ represents a free parameter (in our code units $\kappa$ equals $\frac{3}{2}\Omega_{m} a$ for a cosmological simulation with axion-like dark matter).  Thus, the system can be solved by considering the initial conditions:
\begin{align*}
&\chi(0) = 1, \\
&\chi'(0) = 0, \\
&\chi'''(0) = 0,
\end{align*} 
where $\chi''(0)$ is a free parameter, which is set by requiring asymptotic vanishing solution $\chi(\pm\infty) = 0$. The oscillation period $T$ can then be computed from the resulting solution, for more details see \citet{2014NatPh..10..496S} and \citet{2015MNRAS.451.2479M}.

The soliton profile, normalised such that the total mass is unity, is given by:
\begin{align}
\chi(r) = \frac{32}{\sqrt{33r_{\rm core}^3}\pi}\frac{1}{\left[1+\left(r/r_{\rm core}\right)^2\right]^4},
\end{align}
where $r$ is the distance from the center of the box and:
\begin{align}
m = \sqrt{\frac{33}{r_{\rm core}\kappa}}\frac{32}{\alpha^2},~~~T = \sqrt{\frac{33r_{\rm core}^3}{\kappa}}\frac{\pi^2}{16\gamma^2},
\end{align}
with $\alpha = 0.230$ and $\gamma = -0.692$. In our tests, even if this derivation assumes an infinite box, we use either fixed or periodic boundary conditions. As long as the core radius of the soliton is small enough compared to the simulation box, this solution represents a good approximation.

\end{document}